\documentclass[fleqn,10pt]{wlscirep}
\usepackage[utf8]{inputenc}
\usepackage[T1]{fontenc}
\usepackage{siunitx}
\usepackage{float}
\usepackage{graphicx}
\usepackage{subfig}
\usepackage{cancel}
\usepackage{color,soul}

\newcounter{commentzaehler}

\renewcommand\hl[1]{#1} % disable highlights	
\title{Spatiotemporal patterns of adaptation-induced slow oscillations in a whole-brain model of slow-wave sleep}

\author[1,2,*]{Caglar Cakan}
\author[1,2]{Cristiana Dimulescu}
\author[3]{Liliia Khakimova}
\author[3]{Daniela Obst}
\author[3, 4]{Agnes Flöel}
\author[1,2]{Klaus Obermayer}
\affil[1]{Department of Software Engineering and Theoretical Computer Science, Technische Universität Berlin, Germany}
\affil[2]{Bernstein Center for Computational Neuroscience Berlin, Germany}
\affil[3]{Department of Neurology, University Medicine, Greifswald, Germany}
\affil[4]{German Center for Neurodegenerative Diseases, Greifswald, Germany}

\affil[*]{cakan@ni.tu-berlin.de}


\begin{abstract}
During slow-wave sleep, the brain is in a self-organized regime in which slow oscillations (SOs) between up- and down-states travel across the cortex. While an isolated piece of cortex can produce SOs, the brain-wide propagation of these oscillations are thought to be mediated by the long-range axonal connections. We address the mechanism of how SOs emerge and recruit large parts of the brain using a whole-brain model constructed from empirical connectivity data in which SOs are induced independently in each brain area by a local adaptation mechanism. 
Using an evolutionary optimization approach, good fits to human resting-state fMRI data and sleep EEG data are found at values of the adaptation strength close to a bifurcation where the model produces a balance between local and global SOs with realistic spatiotemporal statistics. Local oscillations are more frequent, last shorter, and have a lower amplitude. Global oscillations spread as waves of silence across the brain, traveling from anterior to posterior regions.
These traveling waves are caused by heterogeneities in the brain network in which the connection strengths between brain areas
determine which areas transition to a down-state first, and thus initiate traveling waves across the cortex.
Our results demonstrate the utility of whole-brain models for explaining the origin of large-scale cortical oscillations and how they are shaped by the connectome. 
\end{abstract}

\begin{document}
	
	\flushbottom
	\maketitle
	
	\thispagestyle{empty}
	
	\section*{Introduction}	
	Slow oscillations (SOs) are a hallmark of slow-wave sleep (SWS), during which neuronal activity slowly (<1 Hz) transitions between \textit{up-states} of sustained firing and \textit{down-states} in which the neurons remain almost completely silent \cite{Steriade1993a, Neske2016}. During SWS, large cortical networks collectively depolarize and hyperpolarize, producing high-amplitude oscillations of the local field potential which can be measured in electroencephalography (EEG) \cite{Massimini2004}. These oscillations play a crucial role for memory consolidation during deep sleep \cite{Diekelmann2010}. 
	
	Intracranial \textit{in-vivo} recordings \cite{Nir2011} of the human brain, as well as EEG \cite{Vyazovskiy2011, Malerba2019}, show that SOs can cover a wide range of participation of brain areas. While the majority of SOs remain locally confined in a few brain regions, some can recruit the entire brain. They preferably originate in anterior parts and propagate to posterior parts of the cortex, like a traveling wave \cite{Massimini2004, Nir2011, Mitra2015, Malerba2019}. \textit{In-vitro} recordings of isolated cortical tissue \cite{Sanchez-Vives2000, Capone2019} demonstrate that SOs can be generated in the absence of any external neural inputs. Taken together, these observations support the idea that SOs are generated locally in an individual brain region, while the synchronized propagation of SOs across the cortex is shaped by the global structure of the human connectome.
	
	While not all details of the cellular processes underlying SOs are known, hyperpolarizing spike-frequency adaptation currents, mediated by activity-dependent potassium concentrations, are thought to play a major role \hl{in terminating the \textit{up-state}} \cite{Sanchez-Vives2000, Neske2016}. 
	This mechanism has been explored in models of isolated cortical masses where activity-dependent adaptation lead to slow oscillations between high-activity \textit{up-states} and a low-activity \textit{down-states} \cite{Tartaglia2017, Nghiem2020, Cakan2020}.
	
	In this paper, we address the question of how \textit{brain-wide} spatiotemporal SO patterns emerge using a large-scale network model of the human brain during SWS. Whole-brain models have been shown to be capable of reproducing functional resting-state patterns of the awake brain \cite{Deco2011, Breakspear2017} from functional magnetic resonance imaging (fMRI) \cite{Deco2009, Cabral2017}, EEG \cite{Endo2020} and MEG \cite{Cabral2014b, Deco2017}. Previous whole-brain modeling studies explaining brain activity during sleep solely focused on fitting models to fMRI functional connectivity. While a pioneering study \cite{Deco2014b} showed that a whole-brain model with adaptation-induced SOs can be fitted to resting-state fMRI, following modeling work did not include SO activity \cite{Jobst2017, Ipina2020} and, thus, did not explore the spatiotemporal activity patterns present during slow-wave sleep on a brain-wide scale.

	To address the emergence of brain-wide SO activity during sleep, we construct a whole-brain model in which cortical brain areas are represented by a mean-field neural mass model \cite{Cakan2020, Augustin2017} with a bistability of \textit{up-} and \textit{down-states}. Adding a local spike-frequency adaptation mechanism to each brain area induces transitions between these states which in turn affect the global brain dynamics. A state-space analysis of the whole-brain model reveals several possible dynamical attractors of the system. To determine optimal model parameters, and hence, the optimal location of the model in state space, we then use a multi-objective evolutionary optimization framework to fit the whole-brain model to resting-state fMRI and sleep EEG recordings simultaneously. This procedure identifies an operating point in which the whole-brain model produces realistic sleep-like SO activity, which we characterize in detail.
	
	Specifically, we find that N3 sleep EEG power spectra are only accurately reproduced if spike-frequency adaptation is included and the adaptation strength is at values close to the boundary of the bifurcation to a state of globally synchronous and self-sustained SOs. %Best fits are found at \comment{critical} values of the adaptation strength parameter, which controls the transitions between up- and down-states in each brain area. 
	Only in this regime, a continuum of locally confined and globally synchronous SOs can be observed, which is consistent with reports of \textit{in-vivo} SO statistics during SWS \cite{Massimini2004, Nir2011, Kim2019, Nghiem2020} and can therefore be considered as a condition for creating a realistic model of whole-brain SOs. In line with these experiments, we also find that local oscillations are shorter, more frequent, and have a lower amplitude, compared to global ones. 
	
	Global oscillations that involve multiple brain areas travel as "waves of silence" across the cortex with a preferred origin of waves in prefrontal areas. The heterogeneous distribution of connection strengths in the human connectome guides the propagation of these waves from anterior to posterior regions. Finally, we explore the role of spike-frequency adaptation of the underlying neurons in affecting brain-wide SO statistics, a mechanism which is subject to cholinergic modulation when the real brain falls asleep \cite{McCormick1989, Nghiem2020}.
	
	Our results demonstrate the utility of whole-brain models for representing a wider range of brain dynamics beyond the resting-state, which highlights their potential for elucidating the origin and the dynamical properties of large-scale brain oscillations in general. %\new{Our approach demonstrates how whole-brain modeling can be used to tighten the gap between theory and experiment by combining our knowledge on the biophysical mechanism of how SOs are generated on the microscopic scale, namely due to potassium-mediated spike-frequency adaptation currents, with macroscopic measurements of brain activity from EEG and fMRI.}
	
	\section*{Results}	
	\begin{figure}[t!]
		\centering
		\includegraphics[width=\linewidth]{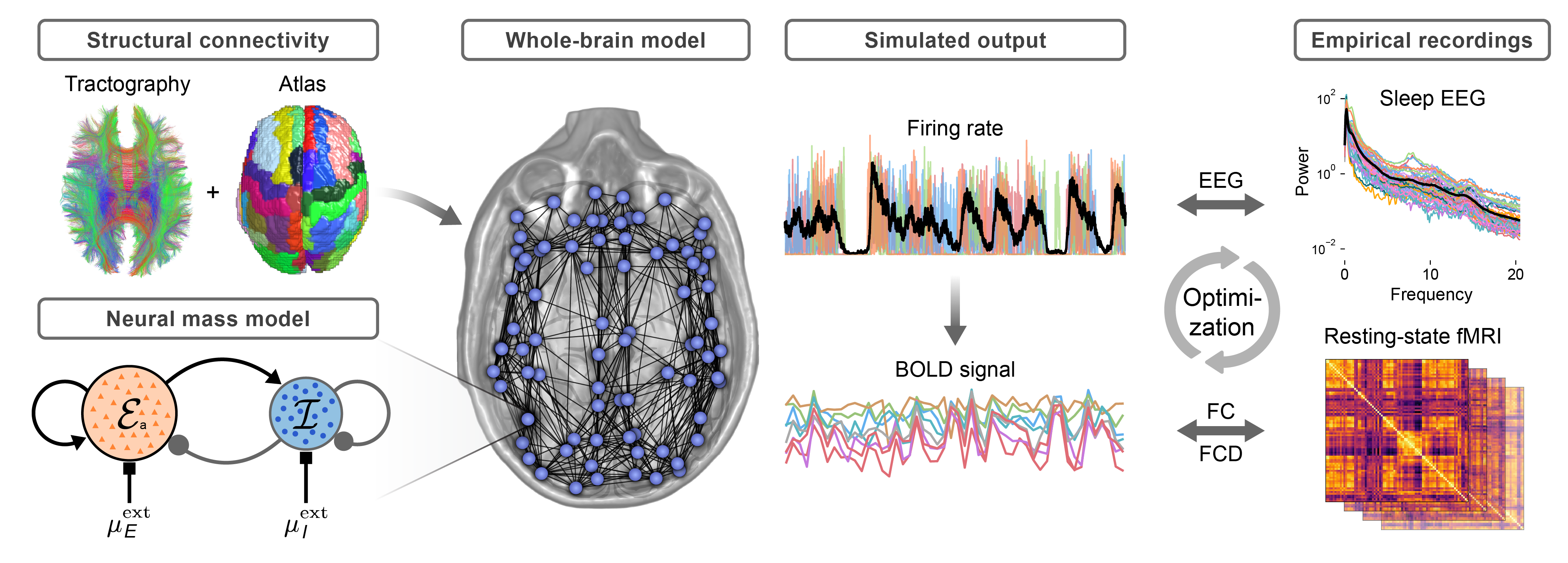}
		\caption{
			\textbf{Construction of the whole-brain model.} Structural connectivity is obtained from probabilistic DTI tractography, and the nodes of the brain network are defined by 80 cortical regions of the AAL2 atlas. The neural mass model represents a single brain area and consists of an excitatory (red) and an inhibitory (blue) population. Long-range connections between areas are via the excitatory populations. All nodes receive a background input currents with means $\mu^\text{ext}_{E, I}$ and noise variance $\sigma_{\text{ou}}$. The excitatory firing rate of each region is converted to a BOLD signal using the hemodynamic Balloon-Windkessel model. For model optimization, the power spectrum of the average excitatory firing rate is compared to the mean EEG power spectrum during sleep stage N3. The functional connectivity (FC) and its temporal dynamics (FCD) are compared to empirical FC and FCD matrices. For details see Methods.
		}
		\label{fig:pipeline}
	\end{figure}
	\subsection*{Neural mass model of a cortical brain region}
	A whole-brain network model is constructed by combining a model of an isolated cortical region with information regarding the structural connectivity of the human brain (Fig. \ref{fig:pipeline}). 
	A mean-field neural mass model \cite{Augustin2017, Cakan2020} derived from populations of excitatory (E) and inhibitory (I) adaptive exponential integrate-and-fire (AdEx) neurons represents a single cortical region as a node in the whole-brain network. The neural mass model has been previously validated against detailed simulations of spiking neural networks \cite{Augustin2017, Cakan2020}. The excitatory subpopulations of every brain area are equipped with an activity-dependent adaptation mechanism.
	
	%\subsection*{Neural mass model}
	%\subsection*{Model description}
	%We construct a mean-field neural mass model of a network of coupled adaptive exponential integrate-and-fire (AdEx) neurons. The neural mass model has been previously validated against detailed simulations of spiking neural networks \cite{Augustin2017, Cakan2020}.
	For a sparsely connected random network of $N \rightarrow \infty$ AdEx neurons, the distribution $p(V)$ of membrane potentials and the mean population firing rate $r_\alpha$ of population $\alpha \in \{E, I\}$ can be calculated using the Fokker-Planck equation \cite{Brunel2000}.
	%Determining the distribution involves solving a partial differential equation, which is computationally demanding. 
	Here, we use a low-dimensional linear-nonlinear cascade model \cite{Ostojic2011, Fourcaud-Trocme2003} of the Fokker-Planck equation which captures its steady-state and transient dynamics via a set of ordinary differential equations and nonlinear transfer functions $\Phi(\mu_{\alpha}, \sigma_{\alpha})$ with the mean membrane current $\mu_\alpha$ and standard deviation $\sigma_\alpha$.
	%For a given mean membrane current $\mu_\alpha$ with standard deviation $\sigma_\alpha$, the mean of the membrane potentials $\bar{V}_\alpha$ as well as the population firing rate $r_\alpha$ in the steady-state can be calculated from the Fokker-Plank equation \cite{Richardson2007} and captured by a set of nonlinear transfer functions $\Phi(\mu_{\alpha}, \sigma_{\alpha})$. 
	%These transfer functions can be precomputed (once) for a specific set of single AdEx neuron parameters (Suppl. Fig. \ref{fig:supp:transfer_functions}).
	
	\subsubsection*{Model equations}
	Every brain area is represented by a node which consists of an excitatory ($E$) and inhibitory ($I$) population. For every node, the dynamics of the mean membrane currents if governed by the differential equation:
	\begin{flalign}
	\label{eq:results_mu_dot}	
	\tau_{\alpha} \, \frac{d{\mu}_\alpha}{dt}  &= \mu^{\text{syn}}_{\alpha}(t)  + \mu_{\alpha}^{\text{ext}}(t) + \mu_{\alpha}^{\text{ou}}(t) - \mu_\alpha(t).
	\end{flalign}	
	Here, $\mu_\alpha$ describes the total mean membrane currents, $\mu^{\text{syn}}_{\alpha}$ the currents induced by synaptic activity from internal connections within a brain area and external connections from other brain areas, $\mu_{\alpha}^{\text{ext}}$ represents the currents from external input sources, and $\mu_{\alpha}^{\text{ou}}$ represents an external noise source which is described by an Ornstein-Uhlenbeck process with standard deviation, i.e. noise strength, $\sigma_{\text{ou}}$, simulated independently for each subpopulation $\alpha$. 
	%= \mu^{\text{syn}}_{\alpha}(K_\text{gl})$ 
	The synaptic currents $\mu^{\text{syn}}_{\alpha}$ depend on the structural connectivity, i.e., the connection strengths and delays between brain regions, and the global coupling strength parameter $K_\text{gl}$ which scales the strengths of all connections globally.
	The input-dependent adaptive timescale $\tau_{\alpha} = \Phi_\tau(\mu_{\alpha})$, the mean membrane potential $\bar{V}_{E} = \Phi_V(\mu_{E})$, and the instantaneous population spike rate $r_\alpha = \Phi_r(\mu_{\alpha})$ of every subpopulation are determined at every time step using precomputed transfer functions (Suppl. Fig. \ref{fig:supp:transfer_functions}). 
	
 	Every excitatory subpopulation is equipped with a somatic spike-frequency adaptation mechanism. The population mean of the adaptation currents $\bar{I}_A$ is \cite{Augustin2017} given by
	\begin{flalign}	
	\label{eq:results_mean_adpatation_current}
	\frac{d{\bar{I}}_A}{dt} &= - \frac{\bar{I}_A}{\tau_A} + b \; \cdot \; r_E(t).
\end{flalign} 	
 	The adaptation currents effectively act like an additional inhibitory membrane current, i.e., $\Phi_{r, \tau, V}(\mu_{\alpha} - \bar{I}_A/C)$. The full set of model equations, including the synaptic equations and the equations for the second statistical moments of all dynamical variables, are provided in detail in the Methods section. 
	%
	%\comment{Parameters $\mu_{E}^{\text{ou}}$, $\mu_{I}^{\text{ou}}$, }
	\subsection*{State-space description of the whole-brain model}
	\begin{figure}[t]
		\centering
		\includegraphics[width=\linewidth]{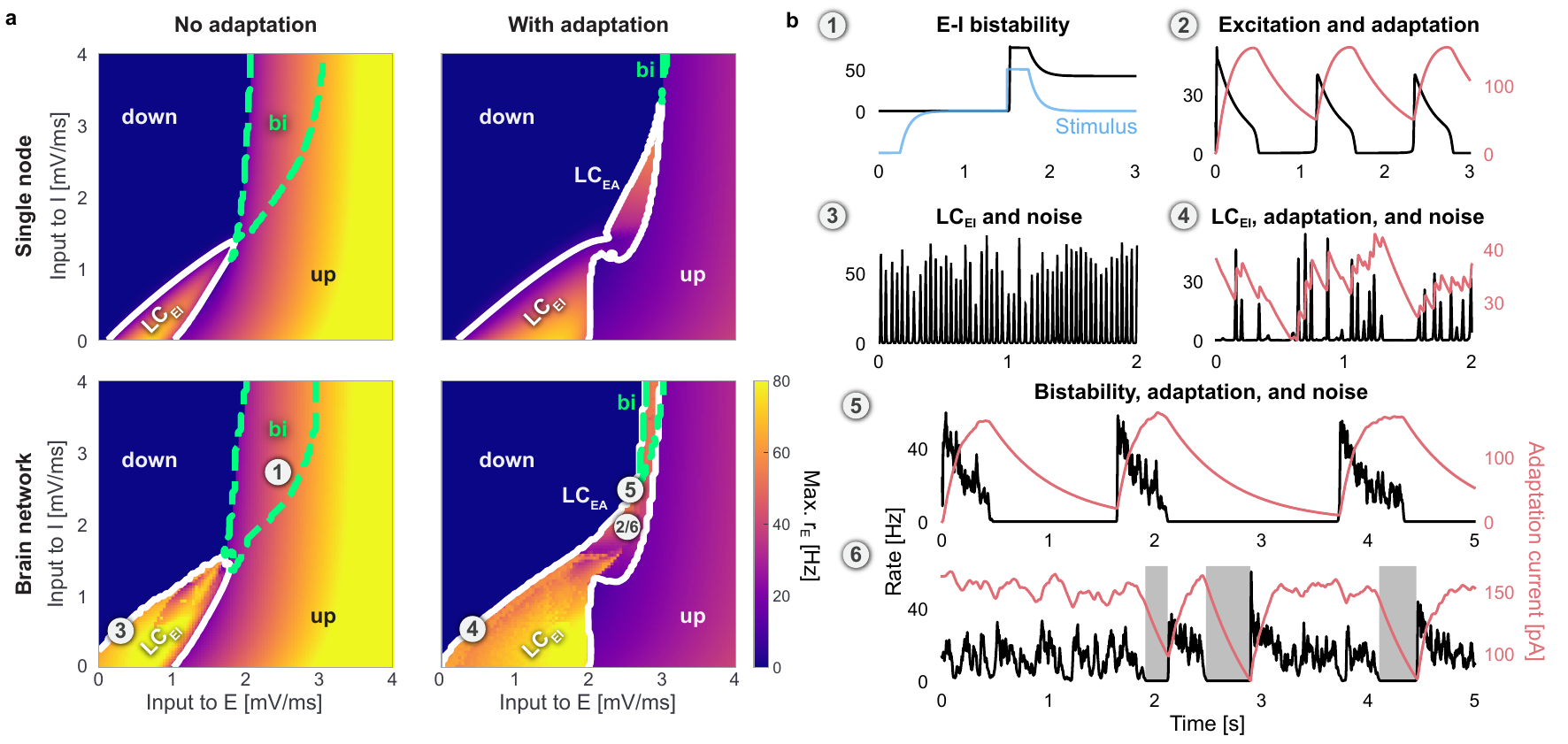}
		\caption{
		\textbf{State space of the brain network.} 
		\textbf{(a)} Single node (top row) and the whole-brain network (bottom row) without ($b=0$ pA, left column) and with spike-frequency adaptation ($b=20$ pA, right column). Horizontal and vertical axes denote the mean input to the excitatory (E), $\mu^\text{ext}_E$, and to the inhibitory (I) populations, $\mu^\text{ext}_I$ . Colors denote the maximum firing rate $r_\text{E}$ of all E populations.
		Regions of low-activity \textit{down-states} (down) and high-activity \textit{up-states} (up) are indicated. Dashed green contours indicate bistable (bi) regions where both states coexist.
		Solid white contours indicate oscillatory states with fast E-I (LC\textsubscript{EI}) and slow excitation-adaptation (LC\textsubscript{EA}) limit cycles. 
%		\comment{Enabling adaptation replaces bi with LC\textsubscript{EA}.}
		\textbf{(b)} Time series of the firing rate $r_\text{E}$ (black) and the adaptation current $I_A$ (red) of one node (left precentral gyrus) in the whole-brain network at several locations in the state space:
		\textbf{(1)} Bistable \textit{up-} and \textit{down-states} are reached through a decaying stimulus (blue) that is delivered to all nodes.
		\textbf{(2)} Finite adaptation causes slow oscillations in the LC\textsubscript{EA} region.
		\textbf{(3)} When external noise is added, fast oscillations already occur outside but close to the LC\textsubscript{EI} region.
		\textbf{(4)} With adaptation, fast oscillations are slowly modulated.
		\textbf{(5)} Noise-induced \textit{down-to-up} transitions with adaptation occur close to the LC\textsubscript{EI} region.
		\textbf{(6)} \textit{Up-to-down} transitions with \textit{down-states} shown as shaded areas.
		Parameters are $K_\text{gl} = 300$, $\tau_\text{A}=600 \si{\milli \second}$, $\sigma_\text{ou} = 0$ mV/$\text{ms}^{3/2}$ in ((a) , (b1), (b2)) and $\sigma_\text{ou} = 0.1$ mV/$\text{ms}^{3/2}$ else. $\mu^\text{ext}_E$, $\mu^\text{ext}_I$ of marked locations are 1: ($2.3, 2.8$), 2: ($2.5, 2.0$), 3: ($0.3, 0.5$), 4: ($0.4, 0.5$), 5: ($2.6, 2.5$), 6: ($2.5, 2.0$) $\si{\milli \volt}/\si{\milli \second}$. All other parameters are given in Table \ref{tab:parameters}.
		} 
		\label{fig:model}
	\end{figure}
	% bifurcation diagrams
	Figure \ref{fig:model} a shows the state space of an isolated E-I node and the coupled whole-brain network in terms of the mean background input currents $\mu_{\alpha}^{\text{ext}}$ to all E and I subpopulations. 
	The state space of the whole-brain model is closely related to the state space of the isolated E-I system. The oscillatory and the bistable states of the brain network are inherited from the isolated nodes and the transitions between states take place at similar locations in parameter space. Due to the heterogeneous connection strengths for different brain regions, in the whole-brain model, transitions between the depicted states happen gradually for one brain region at a time as the input currents to all brain regions are increased simultaneously. However, the range of the input current strengths at which these gradual transitions happen are so small such that when noise is added to the system, we only observe homogeneous states across all areas (see the Methods section for a more detailed discussion).
	
	Without adaptation, the system can occupy several dynamical states, depending on the mean external inputs to E and I: a \textit{down-state} with almost no activity, an \textit{up-state} with a constant high firing rate corresponding to an irregular asynchronous firing state on the microscopic level \cite{Brunel2000, Cakan2020}, a bistable region where these two fixed-point states coexist (Fig. \ref{fig:model} b1), and a fast oscillatory limit-cycle region LC\textsubscript{EI} (Fig. \ref{fig:model} b3) that arises from the coupling of the E and I subpopulations with frequencies between 15-35 Hz. Without noise, oscillations in different brain areas are at near-perfect synchrony with only the inter-areal signal transmission delays and slight differences in oscillation frequency counteracting perfect synchrony.
	%, in which the firing rate oscillates at frequencies between 15-35 Hz, depending on strengths of the external inputs.
	
	\subsubsection*{Adaptation, bistability and slow oscillations}
	The activity-dependent adaptation mechanism in the excitatory subpopulations lead to hyperpolarizing currents that destabilize the high-activity \textit{up-state} in the bistable regime. As we increase the adaptation strength, a Hopf bifurcation \cite{Tartaglia2017, Cakan2020} gradually replaces the bistable regime with a second limit cycle LC\textsubscript{EA} (Fig. \ref{fig:model} b2 and b4) that produces slow oscillations between the (formerly stable) \textit{up-} and the \textit{down-state}. The oscillation frequencies range from around $0.5$ Hz to $2$ Hz, depending on the external inputs and the adaptation parameters. As with the fast limit cycle, when the whole-brain network is parameterized in the slow adaptation limit cycle, without noise, the resulting oscillations are at near-perfect synchrony across all brain regions.
	
	% mechanism of the slow oscillation
	The feedback mechanism leading to this oscillation can be summarized as follows: Without adaptation, in the region of bistability, the \textit{up-} and \textit{down-states} remain stable for an arbitrarily long time. With adaptation, the high firing rate in the \textit{up-state} leads to an increase of the adaptation currents (controlled by the adaptation strength parameter $b$). These inhibitory currents weaken the response of active neurons, which causes the population activity to decay to the \textit{down-state}. In the \textit{down-state}, the activity-driven adaptation currents decrease (with a slow timescale $\tau_\text{A}$) until the recurrent excitation is strong enough to drive the system back to the \textit{up-state}, ultimately causing a slow oscillation between the two originally bistable states. 
	
	% interplay of adaptation parameters
	Therefore, when the system is placed in the bistable region, the interplay of the adaptation strength $b$, the adaptation timescale $\tau_\text{A}$, the strength of mean external inputs, and fluctuations in the form of external noise can lead to stochastic switching between \textit{up-} and \textit{down-states} (Fig. \ref{fig:model} b5 and b6). The resulting pattern of state transitions resembles the bistable neural activity often observed experimentally and in computational models \cite{Sanchez-Vives2000, DAndola2018a, Capone2019, Nghiem2020}. 	
	
	\subsection*{Multi-objective evolutionary optimization}
	\begin{figure}[ht!]
	\centering
	\includegraphics[width=\linewidth]{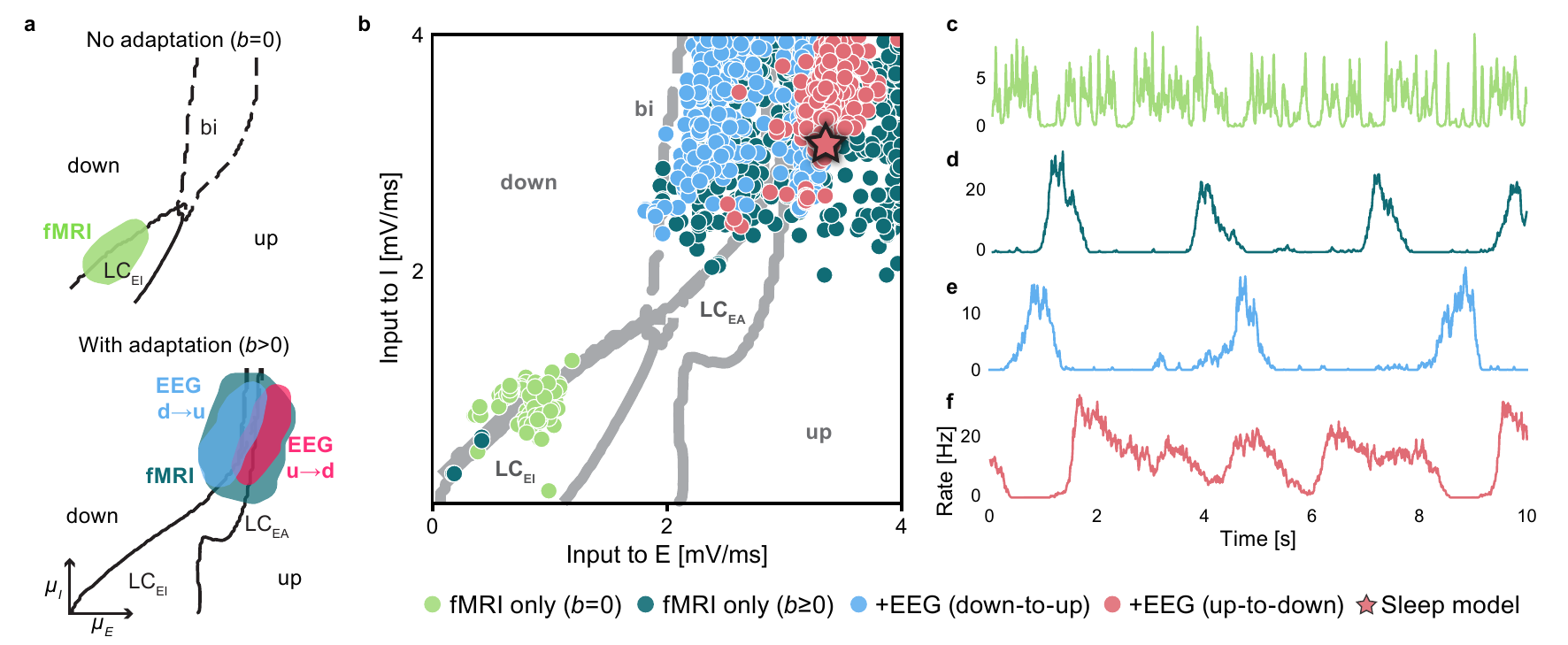}
	\caption{\textbf{Optimization results.} 
		\textbf{(a)} Regions of good fits in the multi-objective optimization. Shown are fMRI fits with FC correlation $> 0.35$ with the empirical data, FCD fits with KS distance $< 0.5$, and EEG fits with a power spectrum correlation $> 0.7$. fMRI fits are shown without adaptation (bright green) and with adaptation (dark green). All EEG fits with \textit{down-to-up} (blue) and \textit{up-to-down} oscillations (red) are shown separately.
		Lines indicate boundaries between the four different dynamical states of the brain network in Fig. \ref{fig:model} a with $b=0 \si{\pico \ampere}$ (top panel) and $20 \si{\pico \ampere}$) (bottom panel).
		\textbf{(b)} Best values of input current parameters to the E and I populations. The colors represent the results from different optimization setups. 
		Good fMRI-only fits appear at the state transition line of the fast limit cycle LC\textsubscript{EI} without adaptation, $b=0 \si{\pico \ampere}$ (bright green) and close to the bistable region with adaptation $b\geq0 \si{\pico \ampere}$ (dark green). Good fMRI+EEG fits are found close to the adaptation limit cycle LC\textsubscript{EA} (blue and red).
		The star symbol indicates the parameters of the sleep model in Fig. \ref{fig:sleep}.
		\textbf{(c-f)} Example average firing rate time series of good fits for all different optimizations. \textbf{(c)} Fit to fMRI data only with $b=0 \si{\pico \ampere}$ 
		\textbf{(d)} and with adaptation $b\geq0 \si{\pico \ampere}$. 
		Fit to fMRI+EEG reveals two classes of solutions \textbf{(e)} with long \textit{down-states} and
		\textbf{(f)} with \textit{in-vivo}-like long \textit{up-states}.
		All parameters are given in Table \ref{tab:parameters}.
	}
	\label{fig:evolution}
	\end{figure}	
	Key parameters such as the mean external inputs $\mu^\text{ext}_E$ and $\mu^\text{ext}_I$, the global coupling strength $K_\text{gl}$, the adaptation strength $b$, the adaptation time scale $\tau_\text{A}$, and the strength of the external noise $\sigma_\text{ou}$ determine the system's dynamics. We perform an evolutionary optimization on all of these parameters simultaneously in order to produce a model with an optimal fit to empirical brain recordings. 
	
	Evolutionary algorithms are stochastic optimization methods that are inspired by the mechanisms of biological evolution and natural selection. 
	The use of a multi-objective optimization method, such as the NSGA-II algorithm \cite{Deb2002}, is crucial in a setting in which a model is fit to multiple independent targets or to data from multiple modalities. In our case, these are features from fMRI and EEG recordings. In a multi-objective setting, not one single solution but a set of solutions can be considered optimal, called the Pareto front, which refers to the set of solutions that cannot be improved in any one of the objectives without diminishing its performance in another one.
	
	For each parameter configuration, a goodness of fit is determined between the simulated model output and the empirical dataset. The simulated BOLD functional connectivity (FC) and the functional connectivity dynamics (FCD) matrices are compared to the empirical fMRI data. The frequency spectrum of the simulated firing rate is compared to EEG power spectra during sleep stage N3. Fitness scores are then averaged across all subjects in the empirical dataset. 	
	
	The optimization is carried out in incremental steps such that the contribution of each step can be assessed individually. 
	The model is first fit to fMRI data only and the optimization is carried out without adaptation ($b=0$) and with adaptation ($b\geq0$, $\tau_\text{A}$ are allowed to vary) separately. The quality of the fitting results to fMRI data is comparable to previous works \cite{Cabral2017, Jobst2017, Demirtas2019} (Suppl. Fig. \ref{fig:supp:fits} b). We then include EEG data as an additional objective in the optimization to derive models with realistic firing rate power spectra. The resulting parameters that produce a good fit to either optimization scheme are shown in Fig. \ref{fig:evolution} b and Suppl. Fig. \ref{fig:supp:fits} a. Details on the optimization procedure are provided in the Methods. 	
	
	\subsubsection*{Fit to empirical data: Up-to-down and down-to-up oscillations}
	Without adaptation, the region of good fMRI fits lies close to the line that marks the transition from a silent \textit{down-state} to the fast E-I limit cycle LC\textsubscript{EI} (Fig. \ref{fig:evolution} a, top panel, and Fig. \ref{fig:evolution} b, bright green dots). The activity in this region shows noisy oscillations with frequencies between 15-35 Hz and brief excursions to the silent \textit{down-state} (Fig. \ref{fig:evolution} c). 
	
	The region of good EEG fits, however, lies in the bistable regime at the boundary to the \textit{down-state}. Here, noise-induced transitions between \textit{up-} and \textit{down-states} produce low-frequency components necessary for a good fit to the empirical power spectrum. 
	Therefore, when no adaptation is present, the regions of good fit to fMRI and EEG are disjoint in state space making it impossible to produce a model that fits to both data modalities simultaneously. 

	With adaptation, however, the bistable region is replaced by a new oscillatory limit cycle LC\textsubscript{EA} around which new regions of good fMRI fits appear (Fig. \ref{fig:evolution} a, bottom panel). In these solutions, the firing rates slowly oscillate between \textit{up-} and \textit{down-states} (Fig. \ref{fig:evolution} d-f). The power spectrum is dominated by this slow oscillation with a $1/f$-like falloff, similar to the empirical EEG power spectrum during SWS (Suppl. Fig. \ref{fig:supp:fit_eeg}). It should be noted that this $1/f$-like falloff is not caused by the noise input to the system alone, since it is not present when the system is parameterized in the fast limit cycle, for example, but originates from the slow transitions between \textit{up-} and \textit{down-states} with a stochastic frequency and state duration.
	
	As a result of the optimization, two classes of good solutions that simultaneously fit well to fMRI and EEG data can be observed: In \textit{down-to-up} solutions (Fig. \ref{fig:evolution} e and Fig. \ref{fig:model} b5), all nodes remain silent for most of the time and the network exhibits short and global bursts of \textit{up-state} activity. In \textit{up-to-down} solutions (Fig. \ref{fig:evolution} f and Fig. \ref{fig:model} b6), however, the activity of the network is in the \textit{up-state} most of the time and slow oscillations are caused by irregular transitions to the \textit{down-state} with a varying degree of synchrony across brain areas. Compared to \textit{down-to-up} solutions, \textit{up-to-down} solutions require stronger external inputs to the excitatory populations (medians $3.36$ vs. $2.73$ \si{\milli \volt / \milli \second}, Fig. \ref{fig:evolution} b), stronger noise $\sigma_\text{ou}$ ($0.36$ vs. $0.27$ \si{\milli \volt / \milli \second^{3/2}}, Suppl. Fig. \ref{fig:supp:fits} a), and a weaker adaptation $b$ ($3.8$ vs. $9.9$ \si{\pico \ampere}). 
	
	\subsection*{Sleep model}
	\begin{figure}[t!]
		\centering
		\includegraphics[width=\linewidth]{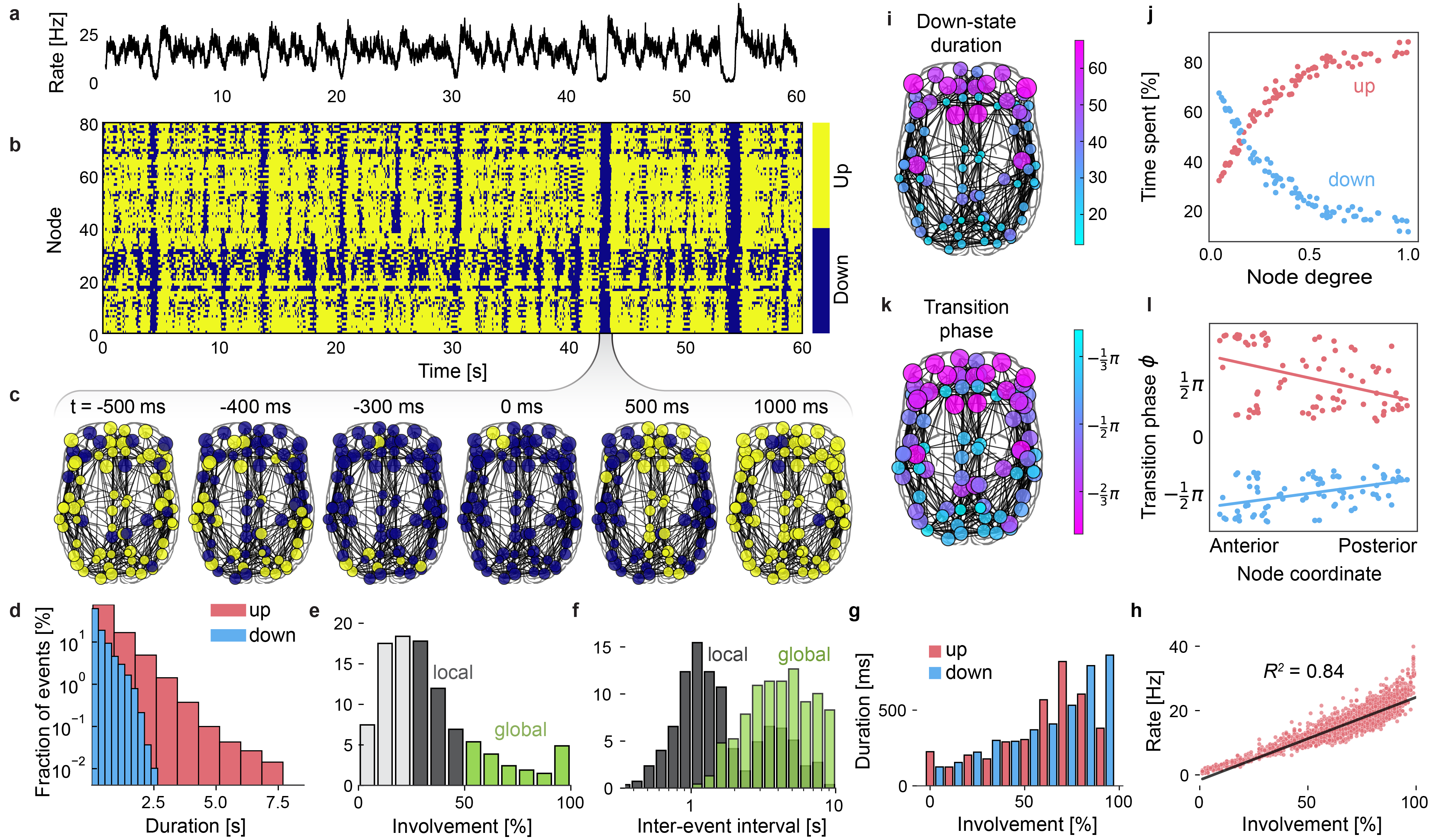}
		\caption{\textbf{Dynamics of the whole-brain sleep model.} 
			\textbf{(a)} Firing rate averaged across all nodes of the whole brain network.
			\textbf{(b)} State time series of all nodes with active \textit{up-} (yellow) and silent \textit{down-states} (blue). 
			\textbf{(c)} Snapshots of state time series plotted on the brain network centered around a global slow wave at t=$43$ s.
			\textbf{(d)} The durations of \textit{up-} (red) and \textit{down-states} (blue) are exponentially distributed.
			\textbf{(e)} Distribution of the \textit{down-state} involvement which measures the fraction of participating brain regions.
			\textbf{(f)} Inter-event interval distributions of local (gray) and global oscillations (green).
			\textbf{(g)} Mean state durations depend on the involvement in \textit{up-} and \textit{down-states}. 
%			\textit{Down-state} durations increase monotonically whereas \textit{up-state} durations show a non-monotonic dependency. 
			\textbf{(h)} The excitatory firing rate depends on the \textit{up-state} involvement of brain regions. Every dot represents a time point of the simulation. Linear regression line is shown in black with regression coefficients $R^2=0.84, p<0.001$. 
			\textbf{(i)} Mean \textit{down-state} duration of areas. Color indicates relative time spent in the \textit{down-state}.
			\textbf{(j)} State durations depend on each brain area's degree. 
			\textbf{(k)} Mean of the \textit{up-to-down} transition phases of all areas (relative to the global oscillation phase $\phi$, see Methods). \textit{Down-states} are preferably initiated in anterior areas.
			\textbf{(l)} \textit{Down-state} transition phases are positively correlated with the anterior-posterior coordinates of brain areas, i.e., they travel from anterior to posterior regions. \textit{Up-state} transitions behave in an opposite manner.
			Parameters are $\mu^\text{ext}_{E} = 3.3 \si{\milli \volt} / \si{\milli \second}$, $\mu^\text{ext}_{I} = 3.7 \si{\milli \volt} / \si{\milli \second}$, $b=3.2 \si{\pico \ampere}$, $\tau_\text{A} = 4765 \si{\milli \second}$, $K_\text{gl}=265$, $\sigma_\text{ou} = 0.37 \si{\milli \volt}/\text{ms}^{3/2}$. All other parameters are given in Table \ref{tab:parameters}.
		} 
		\label{fig:sleep}
	\end{figure}

	Tonic firing with irregular transitions to the \textit{down-state} is indeed typically observed during deep sleep in intracranial \textit{in-vivo} recordings of humans and other mammals \cite{Vyazovskiy2009, Nir2011}. Therefore, we randomly pick one of the parameter configurations from the set of \textit{up-to-down} solutions that resulted from the evolutionary optimization (star in Fig. \ref{fig:evolution} b and Suppl. Fig. \ref{fig:supp:fits} a) as our reference sleep model. To ensure that the properties of the chosen model are typical for all good \textit{up-to-down} fits, we confirmed the results reported below for the 100 best individuals resulting from the evolutionary optimization process (Suppl. Fig. \ref{fig:supp:evolution_criticality}).

	In Fig. \ref{fig:sleep} a, SOs are visible in the average firing rate across all brain areas. In the corresponding node-wise state time series in Fig. \ref{fig:sleep} b, global oscillations that involve the entire brain (cf. Fig. \ref{fig:sleep} c) are visible as vertical lines, and spatially confined oscillations appear as small vertical patches.
	
	\textit{Up-} and \textit{down-state} durations follow a close to exponential distribution with \textit{up-states} having longer durations overall (Fig. \ref{fig:sleep} d), which is in agreement with human \textit{in-vivo} data \cite{Nghiem2020}.
	The involvement of brain areas in \textit{down-state} oscillations, i.e. the proportion of nodes that simultaneously take part in the \textit{down-state}, is skewed towards lower values (Fig. \ref{fig:sleep} e), meaning that most oscillations remain local. Only a small fraction of \textit{down-state} oscillations involve the whole brain, similar to statistics from human intracranial recordings \cite{Nir2011}. The mean \textit{down-state} involvement of all brain regions was $32\%$ with $83\%$ of oscillations involving less than half of the brain.
	Comparing the inter-event intervals (Fig. \ref{fig:sleep} f) of local (e.g. oscillations that involve between $25\%$ and $50\%$ of areas) and global oscillations (involvement above $50\%$), we see that local oscillations happen more frequently than global oscillations, which was also observed experimentally \cite{Kim2019}.
	
	% involvement duration
	\textit{Down-states} last longer, when more brain areas are involved (Fig. \ref{fig:sleep} g), i.e., when areas receive less input on average. Due to the adaptation mechanism of excitatory populations, \textit{up-states} follow a non-monotonic relationship: when more than $65\%$ of areas are involved, excessive excitation leads to a faster increase of adaptation currents which, in turn, shorten \textit{up-states}. 
	The firing rates across all regions also depend on the involvement of brain areas \cite{Nir2011} which means that the more nodes participate, the larger the average firing rate amplitudes of SOs become (Fig. \ref{fig:sleep} h). This relation holds for the average firing rate across all areas, as well the firing rates of individual brain areas, as each of them receives more input when other brain areas as in the \textit{up-state} as well (not shown).
	In summary, due to the interactions of different brain areas, global oscillations are longer, slower, and have a larger amplitude compared to local ones.
	
	% State durations
	In our computational model, the dynamics of SOs and their propagation across the cortex are shaped by the structural properties of the connectome. 
	Since anterior regions have a lower node degree compared to posterior ones (Suppl. Fig. \ref{fig:supp:structural} e), areas which spent the most time in the \textit{down-state} are part of the frontal and the temporal lobe (Fig. \ref{fig:sleep} i and Suppl. Fig. \ref{fig:supp:brain_areas_stats} a). Nodes with a higher degree receive stronger inputs from other areas, lengthening the time spent in \textit{up-states} and shortening \textit{down-states} (Fig. \ref{fig:sleep} j).
	
	\subsubsection*{Traveling waves of silence}
	SOs are known to appear as traveling waves which tend to originate in frontal regions and travel to posterior parts of the brain within a few hundred milliseconds, recruiting multiple brain regions during their propagation \cite{Massimini2004, Nir2011}. Our next goal is to determine, whether our model displays a directionality in the propagation of SOs. Since the only differences between brain areas are due to their connectivity, network properties are expected to play an important role in how oscillations propagate.
	
\begin{figure}[H]
	\centering
	\includegraphics[width=\linewidth]{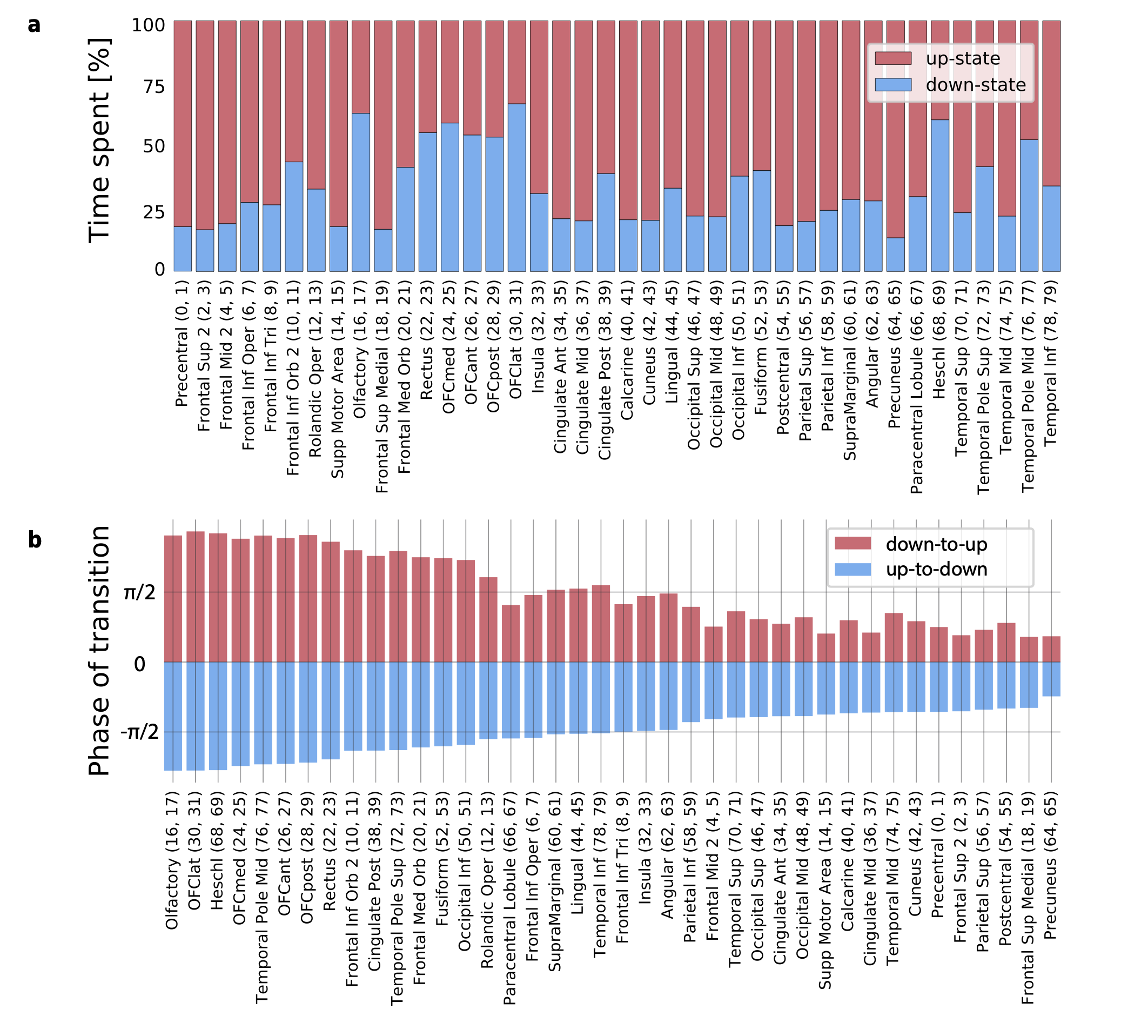}
	\caption{\textbf{State durations and transition phases for the sleep model shown in Fig. \ref{fig:sleep} in the main manuscript.}
		\textbf{(a)} Mean fraction of the duration of each brain area \cite{Rolls2015} spent in the \textit{up-} (red) and the \textit{down-state} (blue). The names and their AAL2 indices of all regions are shown on the x-axis. Durations are averaged across the corresponding contralateral regions. 
		\textbf{(b)} Average transition phase of \textit{down-to-up} transitions (red) and \textit{up-to-down} transitions (blue) for each brain area, sorted by the mean transition phase to the \textit{down-state}. Phases are additionally averaged across contralateral regions.
	}
	\label{fig:supp:brain_areas_stats}
	\end{figure}

	Fig. \ref{fig:sleep} k shows the average timing of the individual \textit{up-to-down} transitions for every brain area with respect to the global \textit{down-state} measured by a transition phase as determined from the involvement time series (see Methods and Suppl. Fig. \ref{fig:supp:sleep_model} a-b). The phases for both up-to-down and down-to-up transitions strongly correlate with the brain areas’ coordinates along the anterior to posterior axis (Fig. \ref{fig:sleep} l).
%	The global whole-brain oscillation phase is determined from the involvement time series (Methods and Suppl. Fig. \ref{fig:supp:sleep_model} a-b) and the mean global phases of all transitions for each brain area (Fig. \ref{fig:sleep} k) are correlated to its coordinate on the anterior to posterior axis of the brain (Fig. \ref{fig:sleep} l).
	\textit{Down-states} tend to appear earlier in the anterior brain and propagate to posterior areas (Fig. \ref{fig:sleep} l, blue line. Linear regression slope $m=3.6 \times 10^{-3}$, $R=0.46, p < 10^{-3}$). Subsequent transitions to the \textit{up-state} happen earlier in posterior and later in anterior areas (Fig. \ref{fig:sleep} l, red line, $m=-6.0 \times 10^{-3}$, $R=-0.41, p < 10^{-3}$). Brain areas that transition to the \textit{down-state} last, tend to initiate \textit{up-states} first (Suppl. Fig. \ref{fig:supp:sleep_model} c), leading to a reversed \textit{down-to-up} wave front propagation. The respective mean transition phases of all regions are shown in Suppl. Fig. \ref{fig:supp:brain_areas_stats} b.
	
	\subsubsection*{Adaptation and noise determine state statistics}
	\begin{figure}[t!]
	\centering
	\includegraphics[width=\linewidth/2]{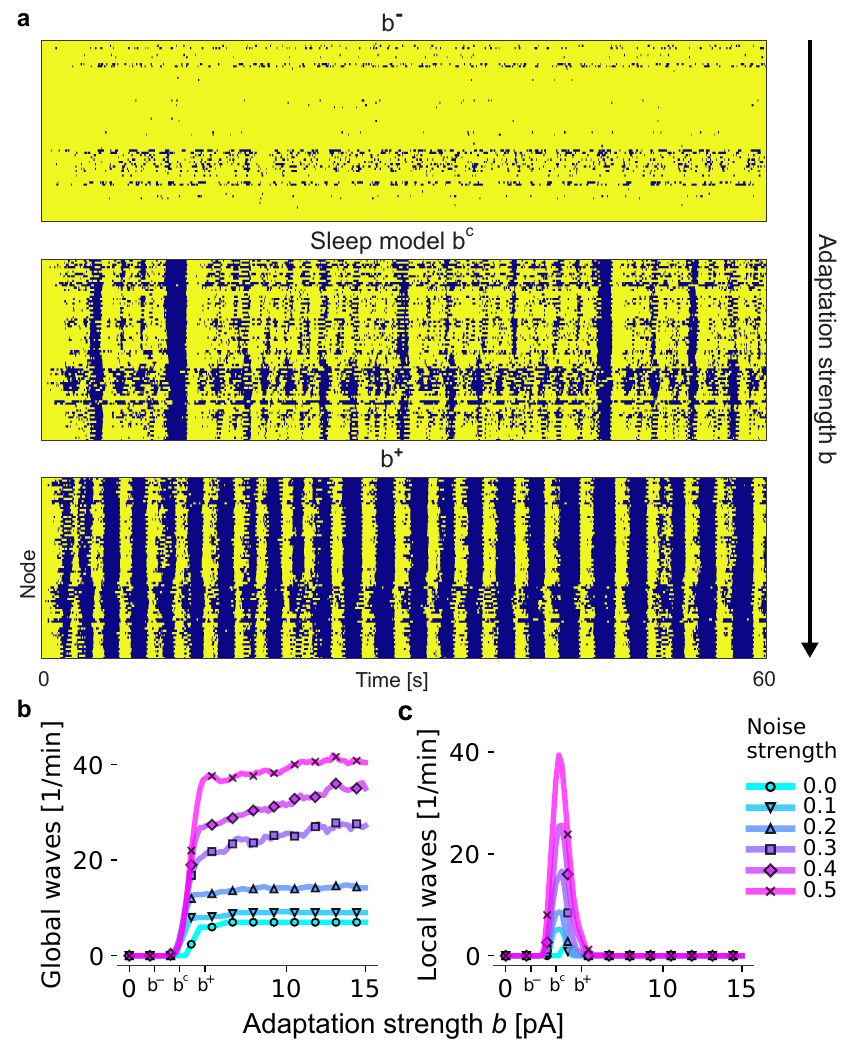}
	\caption{\textbf{Adaptation shapes the spatial pattern of slow-wave activity.} 
		\textbf{(a)} State time series across all nodes in the brain network, with \textit{up-states} in bright yellow and \textit{down-states} in dark blue. Values for the adaptation strength $b$ from the top panel to the bottom panel are $1.6 \si{\pico \ampere}$, $3.2 \si{\pico \ampere}$, and $4.8 \si{\pico \ampere}$. 
		\textbf{(b)} Number of global waves in which at least $50\%$ of all brain regions participate in an SO \textit{down-state} as a function of the adaptation strength $b$ and the input noise strength $\sigma_\text{ou}$ (measured in $\si{\milli \volt}/\text{ms}^{3/2}$, colored lines). Tick marks $b^{-}$, $b^{c}$, and $b^{+}$ indicate values for each panel in (a) respectively.
		\textbf{(c)} Number of local waves per minute with an involvement of $25\%$-$50\%$ of brain areas. 
		All other parameters are as in Fig. \ref{fig:sleep}.
	}
	\label{fig:adaptation}
	\end{figure}	
	The interplay of noise and adaptation, which are both subject to cholinergic modulation in the brain \cite{Nghiem2020}, determines whether SOs remain local or recruit the entire cortex, greatly affecting the SO statistics. 
	% lower b
	Taking the sleep model of Fig. \ref{fig:sleep} with intermediate adaptation strength $b = \SI{3.2}{\pico \ampere}$ as a reference, a reduction of the adaptation strength by $50\%$ ($b = \SI{1.6}{\pico \ampere}$) leads to the disappearance of global oscillations (Fig. \ref{fig:adaptation} a). 
	On the contrary, higher adaptation strengths ($b = \SI{4.8}{\pico \ampere}$) lead to maximum involvement of brain areas for almost every SO. This transition is accompanied by an increase of synchrony across all brain areas (Suppl. Fig. \ref{fig:supp:adaptation} b). 
%	\new{To exclude the possibility that the chosen parameterization of the model has these properties by chance, we conducted the analysis with a set of the 100 best solutions from the evolutionary optimization and can confirm that these effects are also present on the population level (Suppl. Fig. \ref{fig:supp:evolution_criticality}).}
	
	The adaptation strength $b$ acts as a bifurcation parameter, and small adjustments around its threshold value cause a sudden change of the network's dynamics. Figure \ref{fig:adaptation} b shows how the number of global oscillations per minute quickly increases after a critical value $b^{c}$ close to $\SI{3.2}{\pico \ampere}$ is crossed, which is equal to the adaptation strength of the optimized sleep model in Fig. \ref{fig:sleep}. Beyond the threshold, the number of global waves does not significantly increase. For values lower than $b^{c}$, all nodes stay in the \textit{up-state} indefinitely (Suppl. Fig. \ref{fig:supp:adaptation} a). Local waves are observed only around $b^{c}$ (Fig. \ref{fig:adaptation} c), indicating that the system is only able to generate a wide range of local and global oscillations in this regime. Here, the system is in a state of maximum metastability (Suppl. Fig. \ref{fig:supp:adaptation} c) which is an indication for the emergence of complex dynamical patterns. All \textit{up-to-down} solutions with a good fit to the empirical data are close to the threshold value $b^{c}$ (Suppl. Fig. \ref{fig:supp:fits} a).
	
	\section*{Discussion}
	We presented a biophysically grounded human whole-brain model of slow-wave sleep (SWS) that reproduces the observed resting-state fMRI functional connectivity (FC) and its dynamics (FCD), and that captures the EEG frequency spectrum during SWS which is dominated by low frequencies. The model was fitted to multimodal data from fMRI and EEG by the use of a multi-objective evolutionary algorithm \cite{Deb2002}. Good fits to both measurements were only achieved if an activity-dependent adaptation mechanism to the excitatory subpopulations was included. At critical values of the adaptation strength, this resulted in a model in which the interplay of adaptation currents and noise creates a dynamically rich activity with irregular switching between \textit{up-} and \textit{down-states}, similar to the underlying brain activity of slow oscillations (SOs) during SWS \cite{Massimini2004, Nir2011, Nghiem2020}. 
	
	\subsection*{Comparison to human SWS}
	% two classes, in-vivo vs.
	As a result of the optimization procedure, two classes of well-fitting models emerged (Fig. \ref{fig:evolution}), which we named \textit{down-to-up} and \textit{up-to-down} solutions. 
	%In the first class, called \textit{down-to-up} solutions, the activity of all brain areas is in the \textit{down-state} most of the time and is interrupted by short and synchronous transitions to the \textit{up-state}. 
	\textit{Down-to-up} solutions did not produce realistic \textit{in-vivo} SO statistics, since here the simulated brain activity was silent for most of the time. These solutions were more similar to \textit{in-vitro} recordings of SOs \cite{Sanchez-Vives2000, DAndola2018a}
	%, or in some cases to anesthetized animals \cite{Jercog2017, Sanchez-Vives2014, Nghiem2020}, 
	where \textit{down-states} of longer duration than observed during \textit{in-vivo} SWS \cite{Nir2011, Nghiem2020} are interrupted by short bursts of \textit{up-state} activity.
	
	In \textit{up-to-down} solutions, \textit{up-states} were of longer duration and SOs were produced by transitions to \textit{down-states} which represent brief off-periods in neuronal activity. \textit{Up-to-down} solutions require a stronger excitatory input currents, stronger noise fluctuations, and a weaker adaptation strength (Fig. \ref{fig:evolution} b and Suppl. Fig. \ref{fig:supp:fits} a), which all facilitate the initiation of the \textit{up-state} and help sustaining it. 
	Only \textit{up-to-down} solutions reproduced \textit{in-vivo} SO statistics during SWS. Intracranial \textit{in-vivo} data from humans \cite{Nir2011, Nghiem2020} and other mammals \cite{Holcman2006, Vyazovskiy2009} show that \textit{up-states} in the cortex are longer than \textit{down-states}, i.e., that cortical activity is in a tonic and irregular firing state for the most time. 
	
	One explanation for this difference might be that the \textit{up-state} transitions observed \textit{in-vivo} are foremost driven by the convergence of external inputs to brain areas \cite{Chauvette2010, Nir2011} and fluctuations thereof \cite{Jercog2017}, which are both absent in isolated \textit{in-vitro} tissue. This is supported by our modeling results where external mean inputs and noise strengths were key parameters separating both classes.
	
	\subsection*{Balance between local and global oscillations at critical values of the adaptation strength}
	\textit{Up-to-down} transitions can remain local because transitions of individual brain regions have only a weak effect on other brain areas. Global \textit{up-to-down} transitions happen only when the local adaptation feedback currents of a large fraction of brain areas synchronize. \textit{Down-to-up} transitions, however, spread by exciting other brain areas in an all-or-nothing fashion. 
	In fact, most \textit{up-to-down} transitions remain local (Fig. \ref{fig:sleep} e), which is a well-described property of SOs \cite{Nir2011, Vyazovskiy2011, Malerba2019}. 
	
	Global waves last longer, have a higher amplitude (Suppl. Fig. \ref{fig:supp:sleep_model} a and b), and occur less frequently (Fig. \ref{fig:sleep} f) than local waves. 
	In the experimental literature these different characteristics of local and global waves have been used to discern delta waves from SOs \cite{Kim2019}. However, there is still no consensus on whether delta waves and SOs are generated by the same or a qualitatively different underlying neural mechanism \cite{Dang-vu2008}. 
	In the model, local oscillations have these properties because participating regions receive more input from neighboring areas. Here, oscillations with similar characteristics of delta waves do not emerge because of a different generating mechanism, but solely due to the fact that the individual brain areas are embedded in a network and and that the model operates in a regime with a broad involvement distribution.
	
%	 and that the model operates in a critical regime with a broad involvement distribution 
%	 
%	 at which the model is parameterized close but not inside the adaptation-driven low-frequency limit cycle.
%	
%	\subsection*{Critical value of adaptation strength}
	Only at these critical values $b^{c}$ of the adaptation strength parameter where the model is parameterized close but not inside the adaptation-driven low-frequency limit cycle a tight balance between local and global oscillations is maintained (Fig. \ref{fig:adaptation} b and c). The best fits to the empirical data were found in this regime. This optimal working point also coincides with states in which the metastability of the whole-brain dynamics was maximal (Suppl. Fig. \ref{fig:supp:adaptation} c). %\comment{End this paragraph better.}
	
	\subsection*{Slow waves are guided by the connectome}
	In the model, \textit{up-to-down} transitions can originate in many different brain areas and propagate in different directions. However, when averaged over many events, a preferred direction from anterior regions to posterior brain regions becomes evident. This is a well-known feature of SOs \cite{Massimini2004, Nir2011, Mitra2015, Malerba2019} and emerges from the model without any specific adjustments. 
	
	While \textit{down-to-up} transitions spread through excitatory coupling of brain areas, \textit{up-to-down} transitions initiate periods of silence. In the latter case, these oscillations represent a "window of opportunity" in which other brain areas can transition to the \textit{down-state} due to a relative \textit{lack} of inputs. 
	Consistent with prior studies \cite{Hagmann2008}, we measured a positive correlation between the coordinate of a brain region on the anterior-posterior axis and its in-degree (Suppl. Fig. \ref{fig:supp:structural} e). This means that frontal regions receive less input on average and, therefore, spend more time in the \textit{down-state}, as was also observed experimentally \cite{Malerba2019}. As a consequence, \textit{up-to-down} transitions tend to be initiated in frontal areas and spread more easily to other low-degree nodes, ultimately producing wave fronts that travel, on average, from front to back. A mirrored directionality for \textit{down-to-up} transitions (Fig. \ref{fig:sleep} j and Suppl. Fig. \ref{fig:supp:sleep_model} c) can be observed as well which is in agreement with a previous computational study \cite{Roberts2019} that considered cortical waves as spreading activation only. Other heterogeneities that might affect wave propagation, such as a differential sensitivity of brain areas to neuromodulators \cite{Kringelbach2020}, have not been addressed here.
	
	In summary, the increased likelihood of slow waves to be initiated in the prefrontal cortex and the directionality of propagation is determined by the structural properties of the whole-brain model alone. Hence, recognizing that slow waves are propagating as periods of silence explains the possibly unintuitive fact that, although frontal areas are not strongly connected, they become sources of global waves exactly due to their low network degree. 
	
	\subsection*{Underlying neuronal activity of resting-state fMRI is underdetermined}
	Our findings confirm other studies in that best fits to empirical data are found when the system is parameterized close to a bifurcation \cite{Deco2013, Jobst2017, Cabral2017, Demirtas2019} which usually separates an asynchronous firing state from an oscillatory state. In accordance with these studies, we find that low-gamma frequencies (25-35 Hz) can produce good fits to fMRI data.
	However, by including an adaptation mechanism, we also found good fits at the state transition to the slow limit cycle with very low oscillation frequencies (0.5-2 Hz). In some cases, a single \textit{up-state} burst from an otherwise silent brain was enough to produce a good FC fit.
	
	% fc and fcd fits at very different regimes
	This shows that a wide range of parameters and oscillation frequencies can reproduce empirical resting-state (rs) fMRI patterns, primarily because BOLD models act like an infraslow (0.005–0.1 Hz) low-pass filter. The wide range of possible solutions ultimately poses the question which of these regimes is the most suitable candidate for the underlying neuronal activity of rs-fMRI. This problem is particularly evident in sleep. SOs during SWS are profoundly distinct from the tonic firing activity observed in the awake state. Strikingly, however, fMRI FCs during both states are very similar and differ primarily in the average strength of correlations between regions but not significantly in the spatial structure of the FC matrix itself \cite{Dang-vu2008, Mitra2015, Jobst2017}. In agreement with another whole-brain modeling study with adaptive neurons \cite{Deco2014b}, we observed that while the strength of adaptation has a profound effect on the global activity, it did not lead to significantly different FC patterns. Therefore, we think that it is justified to use awake rs-fMRI data for calibrating a whole-brain model of slow-wave sleep, as we do not expect a significant difference compared to fMRI recorded during sleep.
	
	We conclude that it is necessary to incorporate data from faster modalities, such as MEG or EEG, when validating whole-brain models and that the use of fMRI data alone is not sufficient to constrain the vast space of possible models well enough.
	
	\subsection*{Adaptation, cholinergic modulation, and slow oscillations}
	The emergence of SOs caused by activity-dependent adaptation has been thoroughly studied in the past in models of isolated cortical networks \cite{Bazhenov2002, Jercog2017, Cakan2020, Nghiem2020}, and our findings suggest that adaptation plays a major role in the organization of SOs across the whole brain.
	These modeling results are in line with experimental evidence that activity-dependent hyperpolarizing potassium conductances, which contribute to spike-frequency adaptation of pyramidal cells, are in fact responsible for the termination of \textit{up-states} in the cortex \cite{Neske2016}. Furthermore, experiments show that increased levels of acetylcholine (ACh) effectively lead to a weakening of adaptation \cite{McCormick1989}. Analogously, a higher ACh concentration corresponds to a lower adaptation strength $b$ in our model, enabling cholinergic modulation of SO statistics.
	
	The application of carbachol, a drug which causes a blockage of potassium channels and which is an agonist of ACh, can be used to transition a cortical slice from a sleep-like state with SOs to an awake-like state with tonic firing \cite{DAndola2018a}, same as predicted by the model (Suppl. Fig. \ref{fig:supp:adaptation} a).
	Indeed, \textit{in-vitro} experiments have shown that carbachol application leads to a lengthening of \textit{up-states} and a shortening of \textit{down-states} \cite{Nghiem2020}. In line with this, endogenous ACh levels are found to be significantly higher during wakefulness and sharply decrease during SWS \cite{Hasselmo1999} and, vice-versa, that the administration of carbachol enhances transitions from SWS to REM sleep \cite{Carrera-Canas2019}.
	
	In humans, the frequency with which SOs occur changes during a night of sleep and increases with the depth of the sleep stage, ranging from no SOs during the awake stage to around 20 SOs per minute in the deepest sleep stage \cite{Massimini2004}. Similar changes of SO event frequency and durations of \textit{up-} and \textit{down-states} were reported in rats \cite{Vyazovskiy2009}. 
	
	The transitioning of neural dynamics from awake to deep sleep stages as a function of ACh concentrations was previously explored in models of single populations of thalamocortical networks \cite{Bazhenov2002, Hill2005, Krishnan2016}. Combined with the experimental evidence of the significance of cholinergic modulation during sleep, this strongly suggests that the same principles can be applied on the level of the whole brain. In our model, increasing the adaptation strength $b$, which is akin to lowering ACh concentrations in the cortex, also lead to an increase of the number of global oscillations per minute (Fig. \ref{fig:adaptation} b). This highlights the importance of the adaptation mechanism of individual neurons during the transitioning from superficial to deeper sleep stages where SOs are abundant and can involve the whole brain.
	
	In accordance with this, it has been shown that transitions into deeper sleep stages that are accompanied by an increase of SO power also lead to a decrease of delta oscillation power \cite{Rosenblum2001}. A similar transition can be observed in the model where increased adaptation values lead to a replacement of faster local oscillations by slower global oscillations (Fig. \ref{fig:adaptation}).
	
	\subsection*{Outlook}
	Despite the abundance of macroscopic waves in the human cortex \cite{Muller2018}, computational models have only recently been used to study them on a whole-brain scale in order to understand how the connectome shapes these oscillations \cite{Atasoy2016, Robinson2016, Roberts2019}. 
	Our work is a step in this direction, linking the emergence of cortical waves to the underlying neural mechanism that governs adaptation in the human brain during SWS. We have shown how the interplay of local properties of brain areas and the global connectivity of the brain shapes these oscillations. The present study also confirms that whole-brain models can represent wider range of brain activity beyond the resting-state.
	
	Apart from oscillations during SWS, which was the main focus of this paper, cortical waves can be observed in many other scenarios, for example evoked by external stimuli \cite{Stroh2013}, during sensory processing \cite{Davis2020}, and during the propagation of epileptic seizures \cite{Proix2018}. They can range from the mesoscopic \cite{Muller2018} to the whole-brain scale \cite{Burkitt2000, Mitra2015}. 
	Therefore, we have reason to believe that a whole-brain modeling approach, that builds on biophysically grounded and computationally efficient neural mass models, in combination with advanced optimization methods, can help to address questions about the origin and the spatio-temporal patterns of cortical waves in the healthy as well as the diseased human brain.

	\newpage
		
	\section*{Methods}
	\subsection*{Neural-mass model}
	We construct a mean-field neural mass model of a network of coupled adaptive exponential integrate-and-fire (AdEx) neurons. The neural mass model \cite{Augustin2017} has been previously validated against detailed simulations of spiking neural networks \cite{Cakan2020}.
	% optional c/p from main text
	The AdEx model successfully reproduces the sub- and supra-threshold voltage traces of single pyramidal neurons found in cerebral cortex \cite{Jolivet2008, Naud2008} while offering the advantage of having interpretable biophysical parameters. The dimensionality reduction by means of a mean-field approximation provides an increase in simulation speed of about four orders of magnitude over the large spiking neural network while still retaining the same set of biophysical parameters and reproducing all of its dynamical states. 
	
	For a sparsely connected random network of $N \rightarrow \infty$ AdEx neurons, the distribution of membrane potentials $p(V)$ and the mean population firing rate $r_\alpha$ of population $\alpha$ can be calculated using the Fokker-Planck equation in the thermodynamic limit $N \rightarrow \infty$ \cite{Brunel2000}.
	Determining the distribution involves solving a partial differential equation, which is computationally demanding. Instead, a low-dimensional linear-nonlinear cascade model \cite{Ostojic2011, Fourcaud-Trocme2003} captures the steady-state and transient dynamics of a population in form of a set of ordinary differential equations. For a given mean membrane current $\mu_\alpha$ with standard deviation $\sigma_\alpha$, the mean of the membrane potentials $\bar{V}_\alpha$ as well as the population firing rate $r_\alpha$ in the steady-state can be calculated from the Fokker-Planck equation \cite{Richardson2007} and can be captured by a set of simple nonlinear transfer functions $\Phi_{r, \tau, V}(\mu_{\alpha}, \sigma_{\alpha})$. These transfer functions can be precomputed (once) for a specific set of single AdEx neuron parameters (Suppl Fig. \ref{fig:supp:transfer_functions}).	
	
	For the construction of the mean-field model, a set of conditions need to be fulfilled: We assume (1) random connectivity (within and between populations), (2) sparse connectivity \cite{Holmgren2003, Laughlin2003}, but each neuron having a large number of inputs \cite{Destexhe2003} $K$ with $1 \ll K \ll N$, (3) and that each neuron's input can be approximated by a Poisson spike train \cite{Fries2001, Wang2010} where each incoming spike causes a small ($c/J \ll 1$) and quasi-continuous change of the postsynaptic potential (PSP) \cite{Williams2002} (\textit{diffusion approximation}). 
	
	\subsection*{Model equations}
	A detailed mathematical derivation of the model equations was provided before in Refs. \cite{Augustin2017, Cakan2020}. Here, we only present the equations of the linear-nonlinear cascade model which were used to simulate the whole-brain network model. Every brain area is represented by a node consisting of an excitatory ($E$) and inhibitory ($I$) population  $\alpha \in \{E, I\}$. All parameters of the whole-brain model are listed in Table \ref{tab:parameters}. For every node, the following equations govern the dynamics of the membrane currents:
	\begin{flalign}
	\label{eq:mu_dot}	
	\tau_{\alpha} \, \frac{d{\mu}_\alpha}{dt}  &= \mu^{\text{syn}}_{\alpha}(t)  + \mu_{\alpha}^{\text{ext}}(t) + \mu_{\alpha}^{\text{ou}}(t) - \mu_\alpha(t), \\
	\label{eq:synaptic_current}
	\mu^{\text{syn}}_\alpha(t) &= J_{\alpha E} \, \bar{s}_{\alpha E}(t) + J_{\alpha I} \, \bar{s}_{\alpha I}(t), \\
	\label{eq:sigma}
	\sigma^2_{\alpha}(t) &= \sum_{\beta \in \{E, I\}} \frac{2 J_{\alpha \beta}^2 \; \sigma^{2}_{\text{s}, \alpha \beta}(t) \, \tau_{s,\beta} \, \tau_m }{ (1+r_{\alpha \beta}(t)) \, \tau_m + \tau_{s,\beta}} 
	+ \sigma_{\text{ext}, \alpha}^{2}.
	\end{flalign}
	Here, $\mu_\alpha$ describe the total mean membrane currents, $\mu^{\text{syn}}_{\alpha}$ the currents induced by synaptic activity, $\mu_{\alpha}^{\text{ext}}$ the currents from external input sources, and $\mu_{\alpha}^{\text{ou}}$ the external noise input. Means and variances are across all neurons within each population. Note that mean currents measured in units of $\si{\milli \volt}/\si{\milli \second}$ can be expressed in units of $\si{\nano \ampere}$ by multiplying with the membrane capacity $C = 200$ pF, i.e. $1 \si{\milli \volt}/\si{\milli \second} \cdot C = 0.2 \si{\nano \ampere}$. $\sigma^2_\alpha$ is the variance of the membrane currents. The parameters $J_{\alpha \beta}$ determine the maximum synaptic current when all synapses from population $\beta$ to population $\alpha$ are active. The synaptic dynamics is given by:
	\begin{flalign}
	\label{eq:s_mean}
	\frac{d{\bar{s}}_{\alpha \beta}}{dt}  &= \tau_{s, \beta}^{-1} \; \big( \big(1 - \, {\bar{s}}_{\alpha \beta}(t)\big) \cdot r_{\alpha \beta}(t) - {\bar{s}}_{\alpha \beta}(t) \big), \\
	\label{eq:s_sigma}
	\frac{d{\sigma}^{2}_{s, \alpha \beta}}{dt}  &= \tau_{s, \beta}^{-2} \; \big( \big(1 - \bar{s}_{\alpha \beta}(t)\big)^2 \cdot \rho_{\alpha \beta}(t) \, + (\rho_{\alpha \beta}(t) - 2\tau_{s, \beta}(r_{\alpha \beta}(t)+1\big)\big) \, \cdot \sigma^{2}_{s, \alpha \beta}(t) \big).
	\end{flalign}
	Here, $\bar{s}_{\alpha \beta}$ represents the mean of the fraction of all active synapses, which is bounded between $0$ (no active synapses) and $1$ (all synapses active), and $\sigma^2_{s, _\alpha\beta}$ represents its variance.

\begin{figure}[H]
		\centering
		\includegraphics[width=0.9\linewidth]{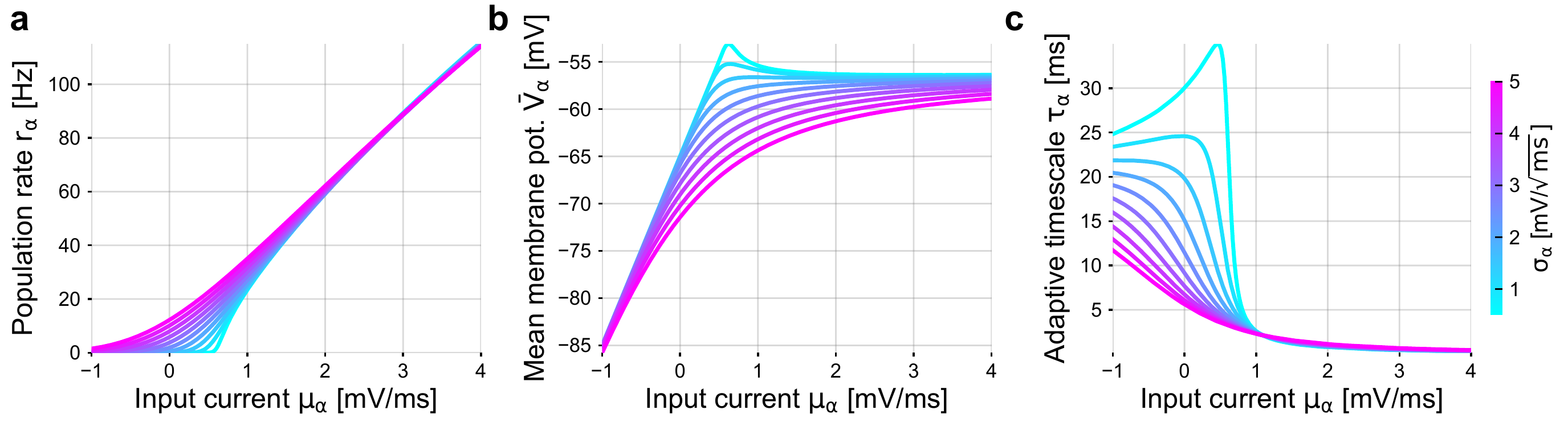}
		\caption{\textbf{Precomputed quantities of the linear-nonlinear cascade model.} 
				\textbf{(a)} Transfer function for the mean population rate.
				\textbf{(b)} Transfer function for the mean membrane voltage.
				\textbf{(c)} Time constant $\tau_{\alpha}$ of the linear filter that approximates the linear rate response function of AdEx neurons. 
				The color scale represents the level of the input current variance $\sigma_{\alpha}$ across the population. All neuronal parameters are given in Table \ref{tab:parameters}.
		}
		\label{fig:supp:transfer_functions}
	\end{figure}

	The input-dependent adaptive timescale $\tau_{\alpha} = \Phi_\tau(\mu_{\alpha}, \sigma_{\alpha})$, the mean membrane potential $\bar{V}_{E} = \Phi_V(\mu_{E}, \sigma_{E})$ and the instantaneous population spike rate $r_\alpha = \Phi_r(\mu_{\alpha}, \sigma_{\alpha})$ are determined at every time step using the precomputed transfer functions (Suppl Fig. \ref{fig:supp:transfer_functions}).
	The mean $r_{\alpha \beta}$ and the variance $\rho_{\alpha \beta}$ of the effective input rate from population $\beta$ to $\alpha$ for a spike transmission delay $d_{\alpha}$ are given by
	
	\begin{flalign}
	r_{\alpha \beta}(t) &= \frac{c_{\alpha \beta}}{J_{\alpha \beta}} \,\tau_{s, \beta} \, \big( K_\beta \cdot r_\beta(t-d_{\alpha}) + \delta_{\alpha \beta E} \cdot
	K_{gl} \sum_{j=0}^{N} C_{ij} \cdot r_\beta(t-D_{ij}) \big) ,\label{eq:mean_rate} \\
	\rho_{\alpha \beta}(t) &= \frac{c_{\alpha \beta}^2}{J_{\alpha \beta}^2} \,\tau_{s, \beta}^2 \, \big( K_\beta \cdot r_\beta(t-d_{\alpha}) + \delta_{\alpha \beta E} \cdot K_{gl} \sum_{j=0}^{N} C_{ij}^2 \cdot r_\beta(t-D_{ij}) \big).
	\label{eq:var_rate}
	\end{flalign}
	$r_\alpha $ is the instantaneous population spike rate, $c_{\alpha \beta}$ defines the amplitude of the post-synaptic current caused by a single spike (at rest, i.e. for ${\bar{s}}_{\alpha \beta} = 0$), and $J_{\alpha \beta}$ sets the maximum membrane current generated when all synapses are active (at ${\bar{s}}_{\alpha \beta} = 1$). $K_{gl}$ is the global coupling strength parameter, $C_{ij}$ are elements from the whole-brain fiber count matrix connecting region $j$ with region $i$, and $D_{ij}$ are elements from the fiber length delay matrix. Here, the Kronecker delta $\delta_{\alpha \beta E}$ is $1$ if $\alpha=\beta=E$ and $0$ otherwise, restricting inter-areal coupling to the excitatory subpopulations only.
	
	\subsubsection*{Adaptation currents}
	For a single AdEx neuron, the hyperpolarizing adaptation current is increased after ever single spike \cite{Benda2003} which leads to a slowly-decreasing spike frequency in response to a constant input \cite{Naud2008}. In the mean-field limit of a large population, an adiabatic approximation can be used to express the mean adaptation current in terms of the mean firing rate of the population. The mean adaptation current is given by $\bar{I}_A$ and acts as an inhibitory membrane current, $\Phi_{r, \tau, V}(\mu_{\alpha} - \bar{I}_A/C, \sigma_{\alpha})$, with its slow dynamics (compared to the membrane time constant) given by \cite{Augustin2017}:
	\begin{flalign}	
		\label{eq:mean_adpatation_current}
		\frac{d{\bar{I}}_A}{dt} &= \tau_A^{-1} \big( a  (\bar{V}_{E}(t)-E_A) - \bar{I}_A \big) + b \; \cdot \; r_E(t).
	\end{flalign}
	We set the subthreshold adaptation parameter $a$ to $0$ and only consider finite spike-triggered adaptation $b$. In Ref. \cite{Cakan2020}, it was shown that a finite $a$ mainly shifts the state space in the positive direction of the excitatory input $\mu_{E}^{\text{ext}}$ and produces no new states compared to when only $b$ is allowed to vary. All parameters are listed in Table \ref{tab:parameters}. 
	
	\begin{table}[t]
	\centering
	\begin{tabular}{lrl}
		\textbf{Parameter} & \textbf{Value} & \textbf{Description} \\
		\hline
		$\mu_{E}^{\text{ext}}$ & $[0 - 4]$ mV/ms & Mean external input to E\\
		$\mu_{I}^{\text{ext}}$ & $[0 - 4]$ mV/ms & Mean external input to I\\
		$\sigma_\text{ou}$ & $[0 - 0.5]$ $\si{\milli \volt}/\text{ms}^{3/2}$ & Noise strength\\		
		$K_e$ & $800$ & Number of excitatory inputs per neuron \\
		$K_i$ & $200$ & Number of inhibitory inputs per neuron \\
		$c_{EE}, c_{IE}$ & $0.3$ mV/ms & Maximum AMPA PSC amplitude \cite{Brunel2003} \\
		$c_{EI}, c_{II}$ & $0.5$ mV/ms & Maximum GABA PSC amplitude \cite{Brunel2003} \\
		$J_{EE}$ & $2.4$ mV/ms & Maximum synaptic current from E to E\\
		$J_{IE}$ & $2.6$ mV/ms &  Maximum synaptic current from E to I\\
		$J_{EI}$ & $-3.3$ mV/ms & Maximum synaptic current from I to E\\
		$J_{II}$ & $-1.6$ mV/ms & Maximum synaptic current from I to I\\
		$\tau_{s,E}$ & $2$ ms & Excitatory synaptic time constant\\
		$\tau_{s,I}$ & $5$ ms &Inhibitory synaptic time constant\\
		$d_{E}$ & $4$ ms & Synaptic delay to excitatory neurons\\
		$d_{I}$ & $2$ ms & Synaptic delay to inhibitory neurons\\
		$C$ & $200$ pF & Membrane capacitance\\
		$g_\text{L}$ & $10$ nS & Leak conductance\\
		$\tau_\text{m}$ & $C/g_\text{L}$ & Membrane time constant\\
		$E_\text{L}$ & $-65$ mV & Leak reversal potential\\
		$\Delta_\text{T}$ & $1.5$ mV & Threshold slope factor\\
		$V_\text{T}$ & $-50$ mV & Threshold voltage\\
		$V_\text{s}$ & $-40$ mV & Spike voltage threshold\\
		$V_\text{r}$ & $-70$ mV & Reset voltage\\
		$T_\text{ref}$ & $1.5$ ms & Refractory time\\
		$\sigma^\text{ext}$ & $1.5$ mV/$\sqrt{\text{ms}}$ & Standard deviation of external input\\
		$E_A$ & $-80$ mV & Adaptation reversal potential\\
		$a$ & $0$ nS & Subthreshold adaptation conductance\\
		$b$ & $[0 - 20]$ pA & Spike-triggered adaptation increment\\
		$\tau_\text{A}$ & $[5 - 5000]$ ms & Adaptation time constant\\
		$K_{gl}$ & $[100 - 400]$ & Global coupling strength\\
		$v_{gl}$ & $20$ m/s & Global signal speed\\
		\hline
	\end{tabular}
	\caption{
		\label{tab:parameters} Summary of model parameters. Parameters which are subject to optimization are given as intervals. 
	}
	\end{table}		
	
	\subsection*{Noise and external input}
	Each subpopulation $\alpha$ of every brain area receives an external input current with the same mean $\mu_{\alpha}^{\text{ext}}(t)$ and standard deviation $\sigma^\text{ext}$ with respect to all neurons of the subpopulation. We set the value of $\mu_{\alpha}^{\text{ext}}(t)$ to draw the state space diagrams in Fig. \ref{fig:model}. The origin of these mean currents is not further specified and can be thought of as a tonic mean background input, for example, from subcortical areas to the cortex. 
	Additionally, every subpopulation also receives an independent noise input $\mu_{\alpha}^{\text{ou}}(t)$ which also originates from unidentified neural sources such as subcortical brain areas and which is modeled as an Ornstein-Uhlenbeck process with zero mean,
	\begin{equation}\label{key}
		\frac{d \mu_{\alpha}^{\text{ou}}}{dt} = - \frac{\mu_{\alpha}^{\text{ou}}}{\tau_{ou}} + \sigma_{\text{ou}} \cdot \xi(t),
	\end{equation}
	where $\xi(t)$ is a white noise process sampled from a normal distribution with zero mean and unitary variance. The noise strength parameter $\sigma_{\text{ou}}$ determines the amplitude of fluctuations around the mean of the process. 
	
	\subsection*{BOLD model}
	The firing rate output $r_E(t)$ of the excitatory population of each brain area is converted to a BOLD signal using the Balloon-Windkessel model \cite{Friston2000, Deco2013} with parameters taken from Ref. \cite{Friston2003}. The BOLD signal is represented by a set of differential equations that model the hemodynamic response of the brain caused by synaptic activity. After simulation, the BOLD signal is subsampled at $\SI{0.5}{\hertz}$ to match the sampling rate of the empirical fMRI recordings.
	
	% move to methods
	%\comment{c/p}
	%The hemodynamic Balloon-Windkessel model \cite{Friston2000} is used to simulate BOLD activity from the firing rate of the E population of each region which is then compared to empirical resting-state fMRI recordings. The average power spectrum of the excitatory firing rates is compared to the EEG power spectra during sleep stage N3 which were obtained from the same subject group (see Methods).	
	
	\subsection*{State space diagrams}
	Due to the semi-analytic nature of the model, the state space diagrams in Fig. \ref{fig:model} a were computed numerically by simulating each point in the diagrams for 20 seconds. Corresponding to the resolution of the diagrams, this resulted in a total of $161 \times 161$ simulations for a single node and $101 \times 101$ simulations for the whole-brain network. State transition lines which mark the bifurcations of the system were drawn by classifying the state of each simulated point and thresholding certain measures to identify abrupt state changes in parameter space. Every simulation was initialized randomly. The state space diagrams were computed without adaptation ($b = 0$) and with spike-triggered adaptation ($b=\SI{20}{\pico\ampere}$). All nodes in the brain network received the same mean background input $\mu^\text{ext}_\alpha$.
	
%	\comment{Mention why the bifurcation diagrams are computed numerically. Note2: I renamed many "bifurcations" to "state transition"}
%	% optional c/p from main text
%	All nodes in the brain network receive the same additional mean background input $\mu^\text{ext}_\alpha$ that is distinct from the input they receive from other nodes in the network. Adding external noise to the mean inputs moves the system along these axes and input fluctuations can cause transitions between dynamical states.
	
	%State transition lines in Fig. \ref{fig:model} a were obtained by $161 \times 161$ simulations for a single node and $101 \times 101$ simulations for the whole-brain network of 20 second simulation time each. All runs had random initial conditions. State space diagrams were computed without adaptation ($b = 0$) and with spike-triggered adaptation ($b=\SI{20}{\pico\ampere}$). 
	
	To classify a state as bistable, a decaying stimulus in negative and subsequently in positive direction was applied to the excitatory population of all nodes. An example of this stimulus can be seen in Fig. \ref{fig:model} b1. This ensured that, if the system was in the bistable regime, it would reach the \textit{down-state} and subsequently the \textit{up-state}. If a difference in the $\SI{2}{\second}$-mean firing rate in any brain area of at least $\SI{10}{\hertz}$ was detected, after the stimulus had relaxed to zero after $\SI{8}{\second}$, it was classified as bistable. A threshold of $\SI{10}{\hertz}$ was chosen because it was less than the smallest difference in firing rate between any \textit{down-} and \textit{up-state} in the state space
	
\begin{figure}[H]
	\centering
	\includegraphics[width=\linewidth]{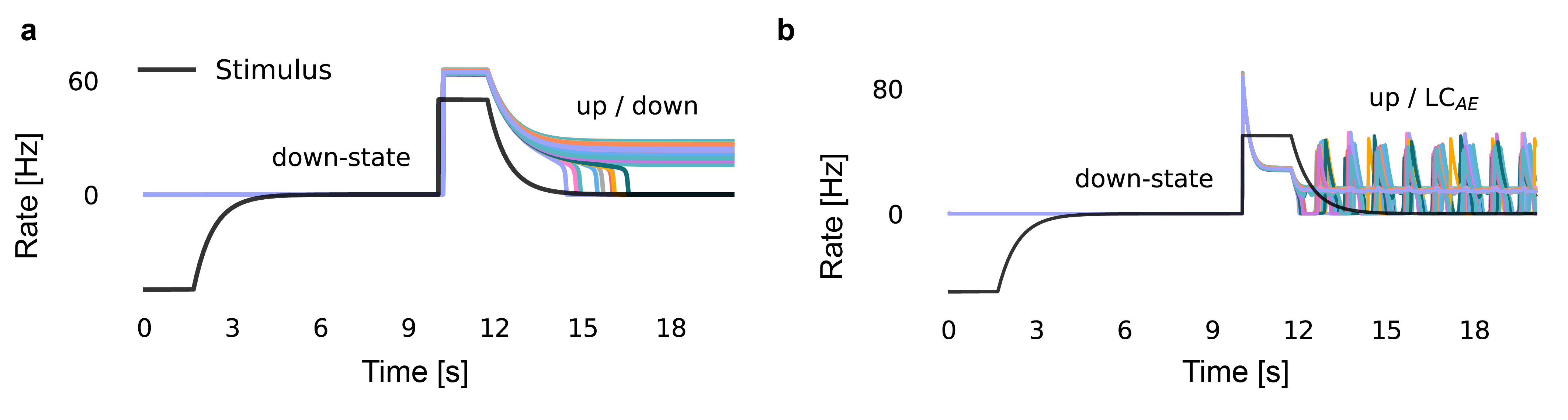}
	\caption{\textbf{Network bistability.}
		\textbf{(a)} The firing rate of each region is shown in different colors. Stimulus (black) pushes the system to the \textit{down-state} and then to the \textit{up-state}. Subsequent relaxation of the stimulus reveals partially bistable states, in which some regions remain in the \textit{up-state} and other regions decay to the \textit{down-state}. Parameters are $\mu^\text{ext}_E = 2.0 \si{\milli \volt / \milli \second}$, $\mu^\text{ext}_I  = 3.5 \si{\milli \volt / \milli \second}$ $\mu_{E}$ and $b = 0 \si{\pico \ampere}$. 
		%, mue = 2.0, mui = 3.5.
		\textbf{(b)} Bistability between the \textit{up-state} and the slow adaptation limit cycle LC\textsubscript{AE}. Parameters are $\mu^\text{ext}_E = 2.8 \si{\milli \volt / \milli \second}$, $\mu^\text{ext}_I  = 3.5 \si{\milli \volt / \milli \second}$ $\mu_{E}$ and $b = 20 \si{\pico \ampere}$. Other parameters are the same as in the state space diagrams in Fig. \ref{fig:model} a in the main manuscript. 
		%b = 20, mue = 2.8, mui = 3.5.
	}
	\label{fig:supp:network_bistability}
	\end{figure}

	% optional c/p from main text
	It should be noted that the state space diagrams of the whole-brain network in Fig. \ref{fig:model} a necessarily provide a reduced description disregarding heterogeneous states that can arise at the depicted state transition boundaries. In the absence of noise, transitions from one network state to another, e.g. from \textit{down-state} to \textit{up-state}, happen gradually as we change the corresponding bifurcation parameter, such as the excitatory external input. Close to the bistable regime some nodes effectively receive more input than others, depending on the in-degree of each node, and can transition to the \textit{up-state} before others do. However, these regions are so narrow that they do not play any meaningful role in the dynamics, i.e., the width of these regions is much smaller than the standard deviation of the noise input. Examples of these states are shown in Fig. \ref{fig:supp:network_bistability} which we could find only by fine-tuning the input strengths of the system without noise. When noise is added, the fluctuating activity of the individual nodes is typically enough to drive the entire network into one of the states in an all-or-nothing fashion.	
	
	Oscillatory regions were classified as such if the oscillation amplitude of the firing rate of at least one brain area was larger than $\SI{10}{\hertz}$ after 8 \si{\second} of stimulation after which all transients vanished. An amplitude threshold of the firing rate oscillations of $\SI{10}{\hertz}$ was chosen because all oscillatory states had a larger amplitude across the entire state space. This was confirmed by choosing a smaller threshold and comparing the transition lines with the former one, which yielded the same results. In the whole-brain network, states were classified as oscillatory, if at least one node was in an oscillatory state. This criterion was compared with a stricter version in which all brain regions had to be simultaneously classified as oscillatory, which again yielded almost the same results with only minimal differences. This leaves us with the conclusion that, in the brain network, the regions in which state transitions from non-oscillatory to oscillatory states happen gradually for each individual brain region, are so narrow that they can be disregarded in our analysis. This is especially true, once noise was added to the system, which made these regions practically unnoticeable.

	\subsection*{Neuroimaging data}
	\subsubsection*{Participants}
	Structural and resting-state functional MRI data were acquired from 27 older adults (15 females: age range = 51-77 years, mean age = 62.93 years; 12 males: age range = 50 - 78 years, mean age = 64.17 years) at the Universit\"atsmedizin Greifswald. For 18 out of the 27 participants, subsequent daytime sleep EEG recordings were also acquired. Nine participants were excluded from the EEG phase due to an inability to sleep in the laboratory. The study was approved by the local ethics committee at Universit\"atsmedizin Greifswald and was in accordance with the Declaration of Helsinki. All participants gave their written informed consent prior to taking part in the study and were reimbursed for their participation.
	\subsubsection*{Structural imaging data}
	\textit{Acquisition.} Data acquisition was performed using a 3T Siemens MAGNETOM Verio syngo B17 MR scanner with a 32-channel head coil. High-resolution anatomical T1 images were acquired using a gradient echo sequence (TR = 1690 ms, TE = 2.52 ms, TI = 900 ms, flip angle (FA) = 9$^{\circ}$, FOV = 250 x 250, matrix size = 246 x 256, slice thickness = 1 mm, number of slices = 176), while for diffusion data a single-shot echo-planar imaging (EPI) sequence (TR = 11100 ms, TE = 107 ms) was used. For every participant, there were 64 gradient directions (b = 1000 s/mm$^{2}$) and one non-diffusion-weighted acquisition (b = 0 s/mm$^{2}$) acquired over a field of view of 230 x 230 x 140 mm, with a slice thickness of 2 mm and no gap, and a voxel size of 1.8 x 1.8 x 2 mm.
	
	\textit{Preprocessing and connectome extraction.} Preprocessing of T1- and diffusion-weighted images was conducted using a semi-automatic pipeline implemented in the FSL toolbox (www.fmrib.ox.ac.uk/fsl, FMRIB, Oxford). Preprocessing of anatomical T1-weighted images involved the removal of non-brain tissue using the brain extraction toolbox (BET) implemented in FSL and the generation of a brain mask. The quality of the brain-extracted images was assessed manually, and, subsequently, 80 cortical regions were defined based on the automatic anatomical labeling (AAL2) atlas introduced in Ref. \cite{Rolls2015}. Brain extraction was also conducted for the diffusion-weighted images, and it was followed by head motion and eddy current distortion correction. Subsequently, a probabilistic diffusion model was fitted to the data using the Bayesian Estimation of Diffusion Parameters Obtained using Sampling Techniques (BEDPOSTX) FSL toolbox. Individual connectomes were obtained by linearly registering each subject's b0 image to the corresponding T1-weighted image, transforming the high-resolution mask volumes from MNI to the individual's diffusion space and running probabilistic tractography with 5000 random seeds per voxel using FSL's PROBTRACKX algorithm \cite{Behrens2007}. Furthermore, as probabilistic tractography contains no directionality information but is dependent on the seeding location, the connection strength $C_{ij}$ between regions \textit{i} and \textit{j} was considered equal to the connection strength $C_{ji}$ between regions \textit{j} and \textit{i} and was obtained by averaging the corresponding entries in the connectivity matrix. Each connectivity matrix was normalized by dividing every matrix entry by the maximum value of the matrix. 
	
\begin{figure}[H]
	\centering
	\includegraphics[width=\linewidth]{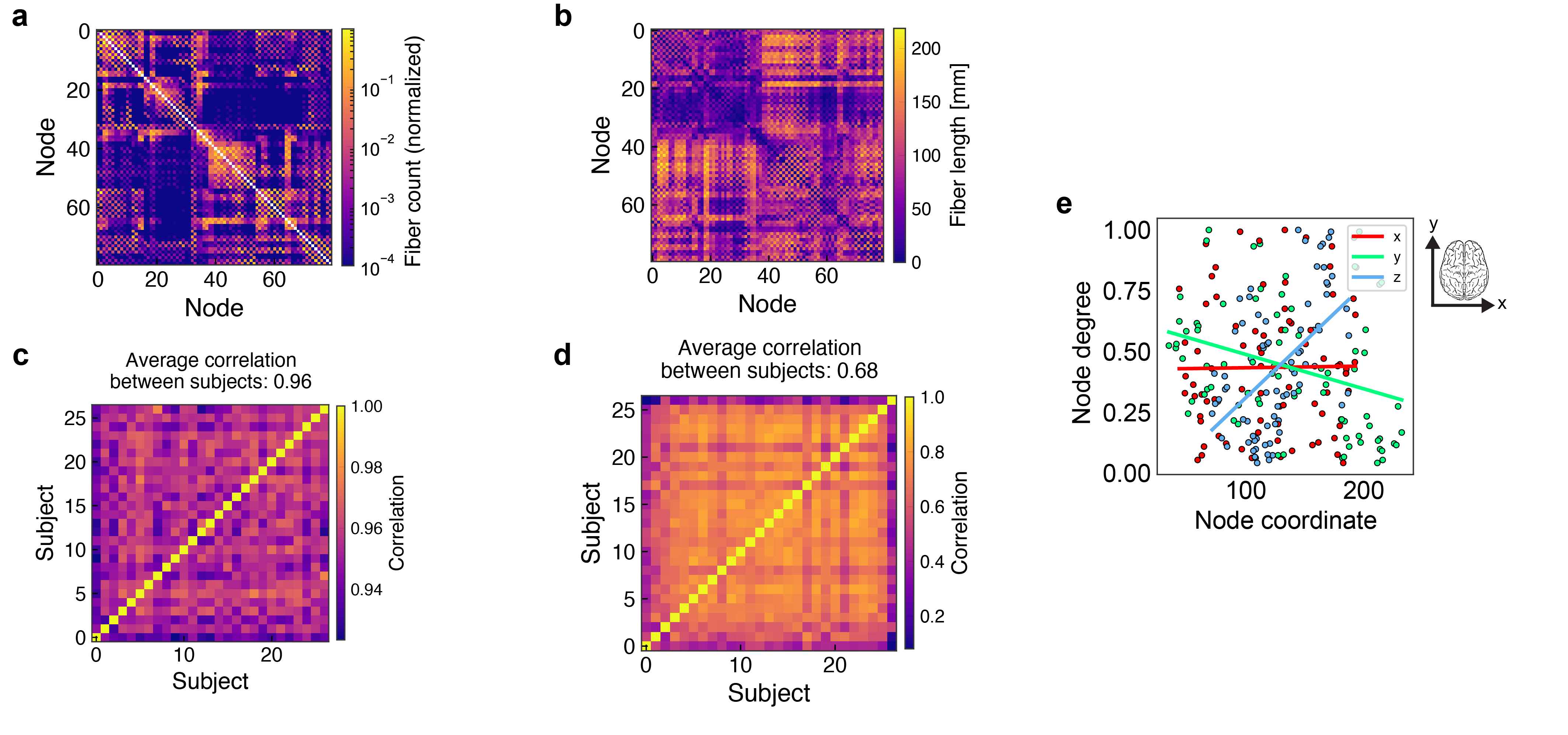}
	\caption{\textbf{Structural brain data.}
		\textbf{(a)} Structural connectivity matrix that defines the connection strengths between cortical regions from the AAL2 atlas. Colors indicate the number of fibers connecting any two regions, normalized by dividing by the maximum fiber count. The matrix is averaged across 27 subjects.
		\textbf{(b)} Fiber length matrix that defines the delay between nodes with the color showing the average length of the fiber bundle between any two regions. The matrix is averaged across all subjects.
		\textbf{(c)} Inter-subject correlation of the lower triangular entries of the structural connectivity matrices. The average inter-subject Pearson correlation of the individual structural connectivity matrices is $0.96$.
		\textbf{(d)} Inter-subject correlation of the fiber length matrices. The average correlation is $0.68$.
		\textbf{(e)} Heterogeneity of node degrees. The weighted node degree of every brain area is plotted against its three spatial coordinates. The x-direction (red) refers to the left-to-right axis of the brain, the y-direction (green) to the posterior-to-anterior, and the z-direction (blue) to the ventral-to-dorsal axis. Coordinates correspond to the center of mass of a brain region. Linear regression lines are shown for each direction. In the x-direction, the linear regression has an insignificant $p$-value such that no dependency can be observed. In the y-direction, the slope is $-1.4 \times 10^{-3}$ with $R^2=0.10$ and $p<0.005$. In the z-direction, the slope is $4.6 \times 10^{-3}$ with $R^2=0.28$ and $p<0.001$.
	}
	\label{fig:supp:structural}
	\end{figure}

	Finally, all subject-specific matrices were averaged to yield a single structural connectivity (Suppl. Fig. \ref{fig:supp:structural} a) and a single fiber length matrix (Suppl. Fig. \ref{fig:supp:structural} b). The resulting connectivity matrix, when multiplied by a coupling parameter $K_{gl}$, determines the coupling strength between any two regions. For a given simulation, the fiber length matrix is divided by the signal propagation speed $v_{gl}$ to yield a time-delay matrix. The average inter-subject Pearson correlation of individual structural connectivity (fiber length) matrices was $0.96$ ($0.68$) (cf. Suppl. Fig. \ref{fig:supp:structural} c, d). 	
	
	\subsubsection*{Resting-state functional MRI data}
	\textit{Acquisition.} Resting-state functional MRI data were acquired using the same 3T Siemens MAGNETOM Verio syngo B17 MR scanner, with the following parameters: TR = 2000 ms, TE = 30, slice thickness = 3 mm, spacing between slices = 3.99 mm, FA = 90$^{\circ}$, matrix size = 64 x 64, FOV = 192 x 192, voxel size = 3 x 3 x 3 mm. Participants were scanned for 12 minutes, leading to an acquisition of 360 volumes per participant.
	
	\textit{Preprocessing and network construction.} Preprocessing of rsfMRI data was conducted using the FSL FEAT toolbox \cite{Woolrich2001}. The first five volumes of each dataset were discarded. Data were corrected for head motion using the FSL McFLIRT algorithm, and high-passed filtered with a filter cutoff of 100 s. Functional images were linearly registered to each subject's anatomical image using FLIRT. A brain mask was created from the mean volume of the data using BET. MELODIC ICA was conducted and artefactual components (including motion, non-neuronal physiological artefacts, scanner artefacts and other nuisance sources) were removed using the ICA FIX FSL toolbox \cite{Salimi-Khorshidi2014}, \cite{Griffanti2014}. Subsequently, the high-resolution mask volumes were transformed from MNI to individual subject functional space and average BOLD time courses for each cortical region were extracted using the \textit{fslmeants} command included in Fslutils.
	\subsubsection*{Sleep EEG data}
	\textit{Acquisition.} Electroencephalography (EEG) recordings were obtained during afternoon naps with a duration of 90 minutes as part of a larger study in which slow oscillatory transcranial direct current stimulation (so-tDCS) was also applied. For the purposes of the current study, however, only the two baseline recordings and one sham (no so-tDCS stimulation) were used. In the baseline sessions, EEG was recorded from 28 scalp sites using Ag/AgCl active ring electrodes placed according to the extended 10-20 international EEG system, while in the sham session, 26 scalp electrodes were used. Data were recorded at a sampling rate of 500 Hz using the Brain Vision Recorder software and referenced to an electrode attached to the nose. Additionally, chin electromiography (EMG) and electrooculagraphy (EOG) data were acquired according to the standard sleep monitoring protocol.
	
	\textit{Preprocessing.} Preprocessing of EEG data was conducted using custom scripts implemented in the FieldTrip toolbox \cite{Oostenveld2011}, in two parts: in the first part, raw data were prepared for an independent component analysis (ICA), which was conducted in order to identify and remove artefactual components, while in the second part, different filtering settings, as required by our main data analysis, were applied again to the raw data and previously identified artefactual independent components (ICs) were removed. In preparation for ICA, data were bandpass-filtered between 1-100 Hz using a finite impulse response filter. In addition, a bandstop filter centered at 50 Hz with a bandwidth of 4 Hz was applied. Subsequently, a manual inspection of the data was conducted in order to remove gross noise artifacts affecting all channels, and ICA was performed using the runica algorithm \cite{Makeig1997}. EMG channels were excluded from the ICA. The resulting 30 maximally independent components (ICs) were visually inspected and those corresponding to muscle artifacts, heart beat, and, where applicable, eye movements were marked for rejection based on scalp topography \cite{Jung1998}, \cite{Jung2000} and power spectrum density \cite{Criswell2010}, \cite{Berry2014}.

	For the main analysis, raw data were bandpass-filtered between 0.1-100 Hz using a finite impulse response filter, then segmented in 10 \si{\second} epochs. Previously identified artefactual ICs were removed from the data, together with the EOG channels. Next, a two-step procedure was employed for detecting remaining artifact-contaminated channels: the first step was based on kurtosis, as well as low- (0.1 - 2 Hz) and high-frequency (30 - 100 Hz) artifacts, while in the second step, the FASTER algorithm was used \cite{Nolan2010}; these channels were afterwards interpolated. In a final step, any 10 \si{\second} epochs still containing artifacts which could not be removed in the previous steps were manually removed from the analysis and the data were linearly detrended.
	
	Sleep stage classification was conducted manually on the raw data, according to the criteria described in Ref. \cite{Rechtschaffen1968}, using the Schlafaus software (Steffen Gais, L\"ubeck, Germany). Here, 30 \si{\second} epochs were used, and each was classified as belonging to one of seven categories: wakefulness, non-REM sleep stage 1, 2, 3, or 4, REM sleep, or movement artifact.

	\subsection*{Model optimization}		
	\subsubsection*{Functional connectivity (FC)}
	From each subject's resting-state fMRI recording, functional connectivity (FC) matrices are obtained by computing the Pearson correlation between the BOLD time series of all brain areas, yielding a symmetric $80 \times 80$ FC matrix per subject. 
	The FC matrix of the simulated BOLD activity (simulation time $T = \SI{12}{\minute}$) is computed the same way. In order to determine the similarity of the simulated and the empirical FC matrices, the Pearson correlation of the lower-triangular elements (omitting the diagonal) between both matrices is computed. For each simulation, this is done for all subjects, and the average of all FC correlation coefficients is taken to determine the overall FC fit. A higher FC correlation coefficient means a better correspondence of simulated and empirical data, with values ranging from -1 to 1 (higher is better). 	
	This ensures that the simulated activity produces realistic spatial correlation patterns.
	
	For the four fitting scenarios, the best subject-averaged FC fit scores were the following: $0.57$ for fMRI-only fits without and $0.56$ with adaptation. With EEG data included in the optimization, fits were $0.57$ for both \textit{up-to-down} and \textit{down-to-up} solutions (Suppl. Fig. \ref{fig:supp:fits} b). For individual subjects, best fits reached up to $0.75$ (Suppl. Fig. \ref{fig:supp:fit_fmri} a and b).
	
	\subsubsection*{Functional connectivity dynamics (FCD)}
	Functional connectivity dynamics (FCD) matrices quantify the cross-correlation between time-dependent FC matrices $FC(t)$. In contrast to computing the grand-average FC alone, this ensures that the temporal dynamics of the simulated and empirical FCs is similar. 
	A rolling window of length $60$ \si{\second} and a step length $10$ \si{\second} is applied to the BOLD time series of length $T = 12$ \si{\minute} to determine $FC(t)$ \cite{Hutchison2013b}. For all $t_1, t_2<T$, the element-wise Pearson correlation of the lower diagonal entries of the matrices $FC(t_1)$ and $FC(t_2)$ is computed, yielding a symmetric $T \times T$ FCD matrix. In order to determine the similarity of the simulated FCD to the empirical FCD, the Kolmogorov-Smirnov (KS) distance of the distributions of lower-triangular elements (omitting the diagonal) of simulated and empirical FCD matrices is calculated. This value ranges from 0 for maximum similarity to 1 for maximum dissimilarity (lower is better). The KS distance is determined for each subject, and an average of all values is taken to determine the overall FCD fit for a simulation. Average FCD fits for all four optimization scenarios were $0.26$, $0.26$, $0.25$, and $0.25$, respectively. The best subject-specific fits reached $0.05$ (Suppl. Fig. \ref{fig:supp:fit_fmri} c and d).
	
	\subsubsection*{Fit to EEG power spectrum}
	%\comment{Einevoll 2013 reports that synaptic currents follow LFP traces the best. Here, we are fitting to the power spectrum of the EEG. We have compared the fits with firing rate to EEG and synaptic current to EEG and have found no significant difference in their power spectra, which is why we use the firing rate for fi}
	Lastly, the frequency spectrum of the firing rate of the excitatory populations of all brain areas is compared to the mean power spectrum of the EEG data during sleep stage N3. 
	%\new{Although the link between local field potentials (LFP), that generate EEG activity, and neural mass models is hard to establish, it was found that the synaptic currents in a network of point neurons as well as their average firing rate (although with a slightly worse performance) can reproduce the LFP time series and its power spectrum \cite{Martinez-Canada2021}. In the present work, we fit the firing rate output of the model to the power spectrum of the EEG. To make sure that our optimization results didn't suffer in fitting performance, we confirmed that the power spectra of the excitatory synaptic activity variable (Eq. \ref{eq:s_mean}) and the firing rate (Eq. \ref{eq:mean_rate}) produce the same power spectrum and, therefore, produce the same optimization results.}
	
	14 out of the 18 subjects with sleep EEG recordings reached the deep sleep stage N3 in a total of 29 sleep sessions. This resulted in a EEG dataset of 1034 non-overlapping epochs of $10$ \si{\second} each. The power spectrum of each epoch was were computed using the implementation of Welch's method \cite{Welch1967} \textit{scipy.signal.welch} in SciPy (v1.4.1) \cite{Virtanen2020}. A rolling Hanning window of length $10$ \si{\second} was used to compute each spectrum. For each channel, the mean of all epoch-wise power spectra was computed. All channel-wise power spectra were then averaged across all channels to yield a subject-specific N3 EEG power spectrum. The power spectra of all subjects were then averaged across all subjects to yield a single empirical EEG power spectrum of N3 sleep. The average and subject-wise EEG power spectra power spectra are shown in Suppl. Fig. \ref{fig:supp:fit_eeg}.
	
	Based on the results of \cite{Martinez-Canada2021}, who showed that in a network of point neurons both the total synaptic current and the average firing rate correlate well with the LFP time series and its power spectrum, we used the firing rate output of the model as a proxy for comparison with the power spectrum of the empirical EEG data. The firing rate of the last $60$ \si{\second} of every simulation was averaged across all $N$ brain regions and the power spectrum was computed using the same method as for the empirical data. 
	We then computed the Pearson correlation coefficient between the simulated and the empirical power spectra between $0$ and $40$ Hz, with correlation values ranging from -1 to 1 (higher is better). During deep sleep, low frequency components (< 1 Hz) are particularly strong due to the underlying slow oscillations (SO) between \textit{up-} and \textit{down-states} \cite{Massimini2004}. Using the power spectrum correlation between the the simulated model and the empirical data, we were able to find model parameters that produce SOs with a frequency spectrum similar to EEG during SWS. Average EEG spectrum correlations for the two scenarios in which the model was fitted to EEG data were $0.93$ when we allowed only \textit{up-to-down} solutions during the optimization, and $0.97$ when \textit{down-to-up} solutions were allowed as well. In addition, we confirmed that the power spectra of the excitatory synaptic activity variable (Eq. \ref{eq:s_mean}) and the firing rate (Eq. \ref{eq:mean_rate}) were similar (Pearson correlation $>0.99$) leading to the same optimization results.
	
	\subsubsection*{Evolutionary algorithm}
\begin{figure}[H]
		\centering
		\includegraphics[width=\linewidth/2]{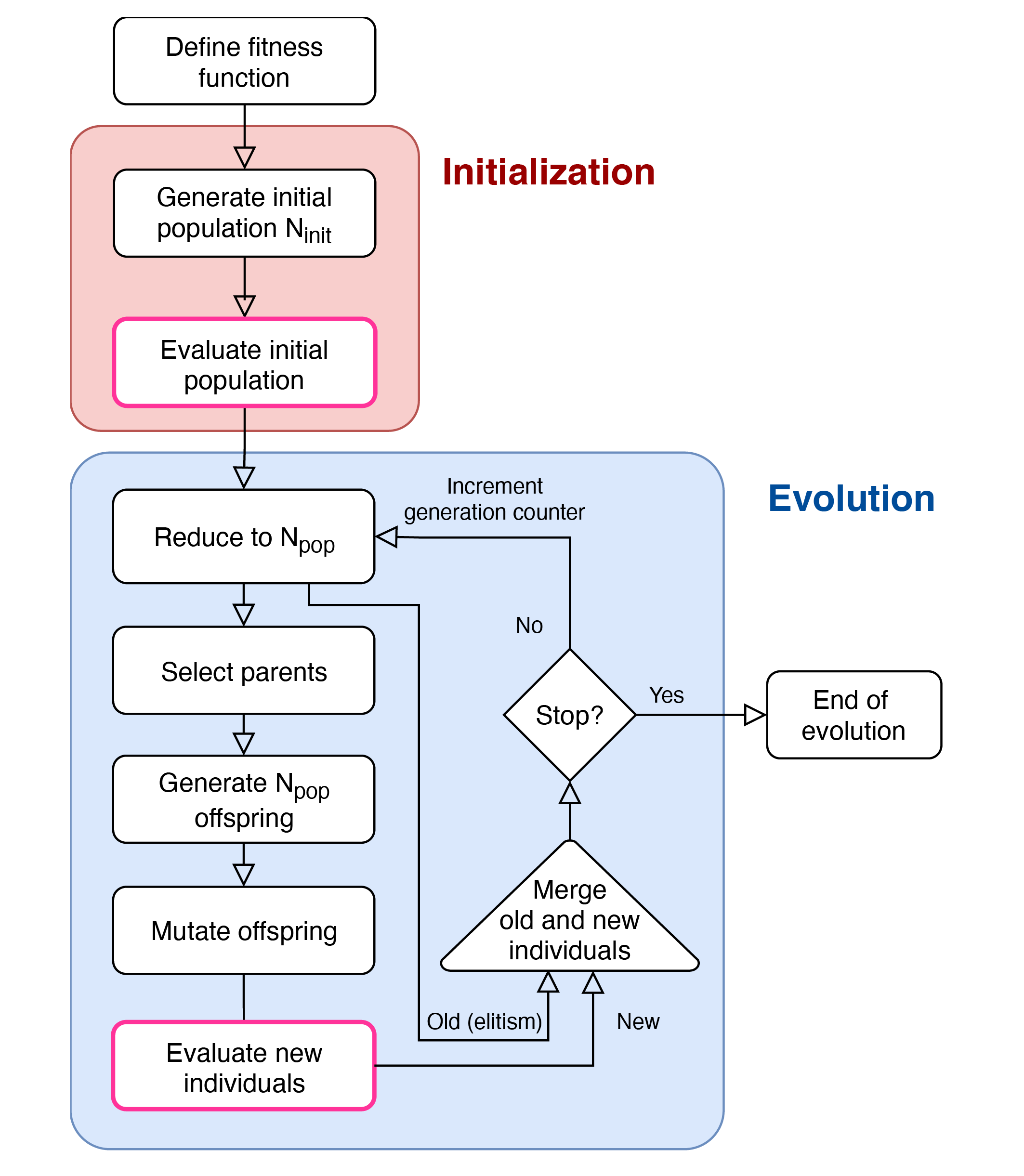}
		\caption{\textbf{Schematic of the evolutionary algorithm}
			The evolution is has two phases with its initialization phase shown in red and its second phase with repeating evolutionary rounds shown in blue. The evolutionary optimization stops when the generation counter reaches a predefined number of generations. The whole-brain model is simulated, and a fitness score is assigned in the pink boxes. $N_\text{init}$ denotes the size of the initial population, and  $N_\text{pop}$ the size of the ongoing population. Details on the algorithm and the evolutionary operators used are provided in the Methods. 
		}
		\label{fig:supp:algorithm}
	\end{figure}

	An evolutionary algorithm is used to find suitable parameters for a brain network model that fits to empirical resting-state fMRI and sleep EEG recordings.
	A schematic of the evolutionary algorithm is shown in Suppl. Fig. \ref{fig:supp:algorithm}. The use of a multi-objective optimization method, such as the NSGA-II algorithm \cite{Deb2002}, is crucial in a setting in which a model is fit to \hl{multiple independent targets} or data from multiple modalities. Algorithms designed to optimize for a single objective usually rely on careful adjustment of the weights of each objective to the overall cost function. In a multi-objective setting, not one single solution but a set of solutions, called the first Pareto front, can be considered optimal. This refers to the set of solutions that cannot be improved in any direction of the multi-objective cost function without diminishing its performance in another direction. These solutions are also called non-dominated.
	
	In the evolutionary framework, a single simulation is called an individual and its particular set of parameters are called its genes, which are represented as a six-dimensional vector, i.e. one element for each free parameter (listed below). A set of individuals is called a population. For every evolutionary round, also called a generation, the fitness of every new individual is determined by simulating the individual and computing the similarity of its output to the empirical data, resulting in a three-dimensional fitness vector with FC, FCD, and EEG fits, determined as described above. Then, a subset of individuals is selected as parents from which a new set of offspring are generated. Finally, these offspring are mutated, added to the total population, and the procedure is repeated until a stopping condition is reached, such as reaching a maximum number of generations. The multi-objective optimization is based on non-dominated sorting and several other evolutionary operators as introduced in Ref. \cite{Deb2002} which are implemented in our software package \textit{neurolib} \cite{cakan2021a} using the evolutionary algorithm framework DEAP \cite{Fortin2012}.
	
	The evolutionary algorithm consists of two blocks, the initialization block and the evolution block (Suppl. Fig. \ref{fig:supp:algorithm}). For initialization, a random population of $N_\text{init} = 320$ (in the fMRI-only case) or $N_\text{init} = 640$ (in the fMRI and EEG case) individuals is generated from a uniform distribution across the following intervals of the model parameters: $\mu^\text{ext}_{E}$ and $\mu^\text{ext}_{I}$ $\in [0.0, 4.0]$ \si{mV}/\si{ms}, $b \in [0.0, 20.0]$ \si{\pico \ampere}, $\tau_\text{A} \in [0.0, 5000.0]$ \si{ms}, $K_\text{gl} \in [100, 400]$, and $\sigma_{\text{ou}} \in [0.0, 0.5]$ mV/$\text{ms}^{3/2}$. Since the parameter space of good fits in the fMRI and EEG case is smaller compared to the fMRI-only case (see Fig. \ref{fig:evolution} a), a larger initial population was used to ensure that the algorithm is able to find these regions. The initial population is simulated, and the fitness scores are evaluated for all individuals.

	We then start the evolutionary block which will repeat until the stopping condition of a maximum number of $20$ (fMRI-only) or $50$ (fMRI and EEG) generations is reached. Using a non-dominated sorting operator, the initial population is reduced to the population size $N_\text{pop} = 80$ (fMRI-only) or $N_\text{pop} = 160$ (fMRI and EEG). From this population, using a tournament selection operator based on dominance and crowding distance, a set of parents is chosen. A simulated binary crossover is used as a mating operator and is applied on the parent population to generate $N_\text{pop}$ new offspring. Finally, a polynomial mutation operator is applied on the offspring population, which introduces randomness and thus aides the exploration of the parameter space. After all offspring have been evaluated and a fitness is assigned to each of them, the population is merged with the parent population. The process is then repeated for each generation such that the algorithm produces improving fits in every new generation. 
	
	The evolutionary process was significantly accelerated by avoiding long ($12$ \si{\minute}) simulations with almost no activity. This was achieved by a stage-wise simulation scheme. In the first stage, every run was simulated for $10$ \si{\second} with a transient time of $1$ \si{\second}. If the maximum firing rate of any brain area did not exceed $10$ \si{\hertz}, the run was omitted and marked as invalid. Then, in the second stage, each valid run was simulated for the full length of $12$ \si{\minute} and the fitness of that run was evaluated.
	
	When fitting to fMRI and EEG data simultaneously, we filtered for \textit{up-to-down} solutions by thresholding the median firing rate in the first stage of each simulation. If the median firing rate (across all nodes and time) was below $1$ \si{\hertz}, the simulation was omitted and marked as invalid to avoid finding \textit{down-to-up} solutions. If the median firing rate was above $15$ \si{\hertz}, the simulation was omitted, to avoid solutions that showed excessive \textit{up-state} activity. No filtering was necessary when the model was optimized for \textit{down-to-up} solutions, since the algorithm had a strong tendency to find these solutions without any intervention. The second stage was not subject to such filtering.		
	
	\subsection*{Sleep model analysis}
	\subsubsection*{Up- and down-state detection}
	\textit{Up-} and \textit{down-states} are detected by thresholding the excitatory firing rate $r_E(t)$ of each brain region, similarly as in Refs. \cite{Renart2010, Nghiem2020}. At any time $t$, a region is considered to be in the \textit{up-state} if $r_E(t)>\theta \cdot max(r_E(t))$ with $\theta = 0.01$, otherwise it is considered to be in the \textit{down-state}. States that were shorter than $\SI{50}{\milli \second}$ were discarded by replacing them with the preceding state. For robustness, the statistics in Fig. \ref{fig:sleep} d-l were computed from $10$ \si{\minute} simulations. To compute the statistics shown in Fig. \ref{fig:adaptation} b and c and Suppl. Fig. \ref{fig:supp:adaptation}, each parameter value was simulated $10$ times for $1$ minute each and results were averaged across the $10$ simulations.
	
	\subsubsection*{Involvement}
	When not stated differently, we report the involvement in the \textit{down-state} because its onset is usually considered to be the beginning of a slow oscillation. Following the definition in Ref. \cite{Nir2011}, the involvement time series $I(t)$ represents a fraction and is defined as the number $n$ of brain areas in a given state at time $t$ divided by the number $N$ of all brain areas: $I(t) = n(t)/N$. Note that maximum involvement in the \textit{down-state} means minimum involvement in the \textit{up-state} and \textit{vice versa}. 
	
	\subsubsection*{Global and local waves}
	Oscillations in Figs. \ref{fig:sleep} f, \ref{fig:adaptation}, and Suppl. Fig. \ref{fig:supp:adaptation} were detected using the peak finding algorithm \textit{scipy.signal.find\_peaks} implemented in SciPy applied on the involvement time series $I(t)$ which ranged between $0-100\%$. For peak detection, the minimum peak height was $10\%$ and the minimum distance to a neighboring peak was of $100 \si{\milli \second}$. A Gaussian filter \textit{scipy.ndimage.gaussian\_filter1d} with a width of $200 \si{\milli \second}$ was applied on $I(t)$ before peak detection. Oscillations were considered global if the amplitude of the peak in $I(t)$ was larger than $50\%$ (i.e. most brain areas participated) and as local if $25\% < I(t) < 50\%$. Oscillations with an involvement of less than $25\%$ were not considered. It should be noted that the choice of these threshold values for the classification into local and global waves is in line with previously used values \cite{Nir2011}.
	
	\subsubsection*{Population statistics}
\begin{figure}[H]
	\centering
	\includegraphics[width=0.8\linewidth]{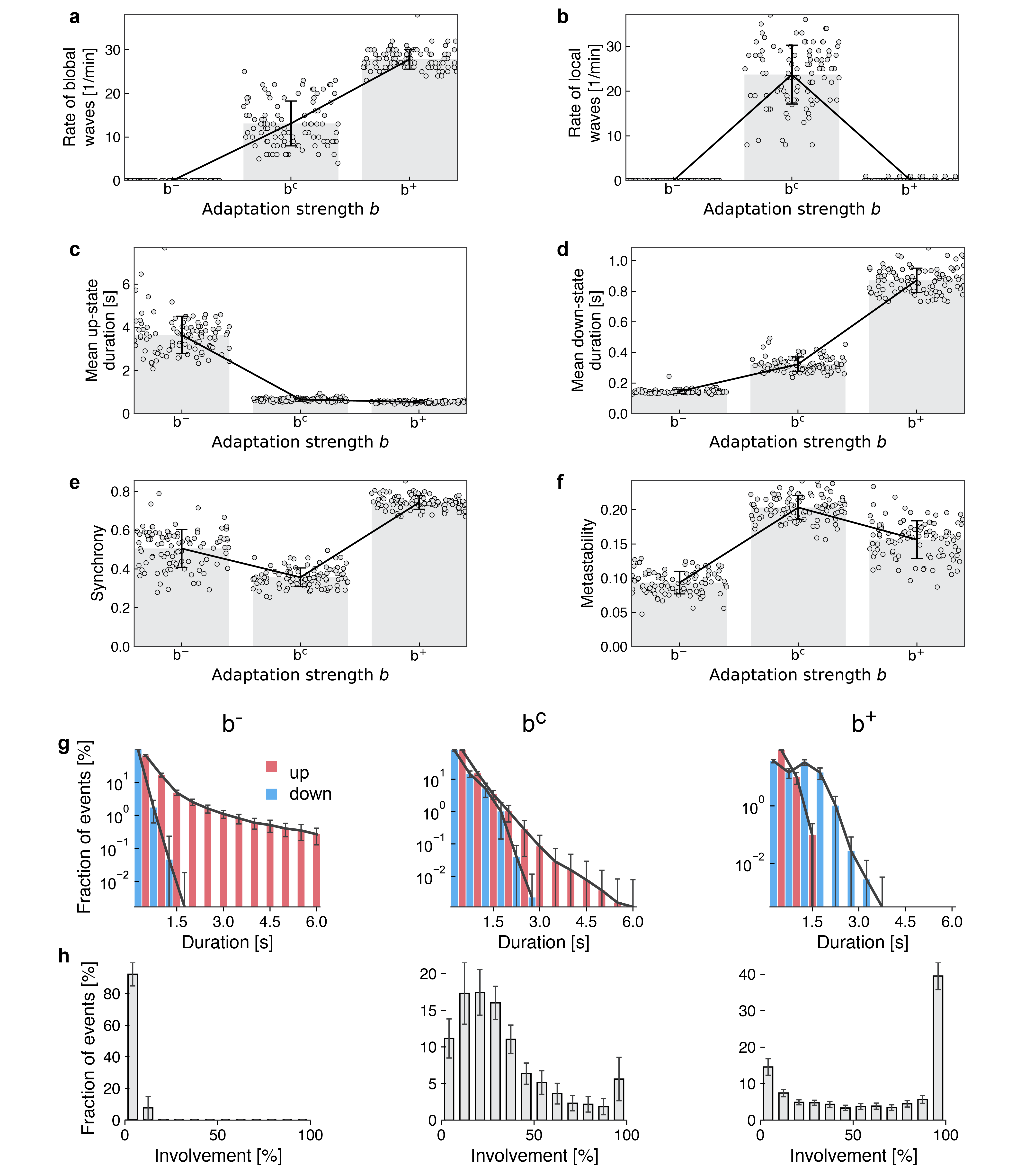}
	\caption{\textbf{Population analysis of adaptation-dependent statistics.}
		Analysis of the “up-to-down” population of sleep models shown in Fig. \ref{fig:evolution} b (red color), which were simultaneously fitted to fMRI and EEG data. The 100 best fitting individuals were analyzed (see Methods) for three different adaptation strengths: the their original (optimized value of the) adaptation strength parameter ($b^{c}$), a $50\%$ reduction ($b^{-}$), and a $50\%$ increase ($b^{+}$), resulting in a total of 300 individuals.
		The panels show: \textbf{(a, b)} The number of global (> 50% involvement) and local (25-50% involvement) waves per minute,
		\textbf{(c,d)} the mean duration of up-states and down-states,
		\textbf{(e,f)} the mean of synchrony and metastability, as measured by the Kuramoto order parameter,
		\textbf{(g)} the distribution of \textit{up-state} (red) and \textit{down-state} (blue) mean durations across time and brain regions (error bars indicate standard deviations) for three levels of adaptation strength $b$, and 
		\textbf{(h)} the distribution of mean \textit{down-state} involvement (error bars indicate standard deviations). 
		All other parameters are as in Fig. \ref{fig:sleep} in the main manuscript.
	}
	\label{fig:supp:evolution_criticality}
	\end{figure}

	To ensure that the properties of the chosen model are typical for all good \textit{up-to-down} fits, we confirmed the results reported in Fig. \ref{fig:sleep} for the 100 best individuals resulting from the evolutionary optimization process (Suppl. Fig. \ref{fig:supp:evolution_criticality}).
	First, the fMRI+EEG (up-to-down) population shown in Fig. \ref{fig:evolution} b, was sorted by the score of the individuals which is the weighted sum of each individual's fitness. Then, the best 100 individuals were selected, and each individual was replicated twice. For each replica, the adaptation strength parameter $b$, which was subject to optimization and had a value of $b^{c}$, was increased ($b^{+}$) and decreased ($b^{-}$) by $50\%$. The resulting 200 new individuals were then simulated for 1 \si{\minute} each and the statistics were obtained for all 300 individuals, similar to the sleep model presented. 

	\subsubsection*{Whole-brain oscillation phase}
\begin{figure}[H]
	\centering
	\includegraphics[width=\linewidth]{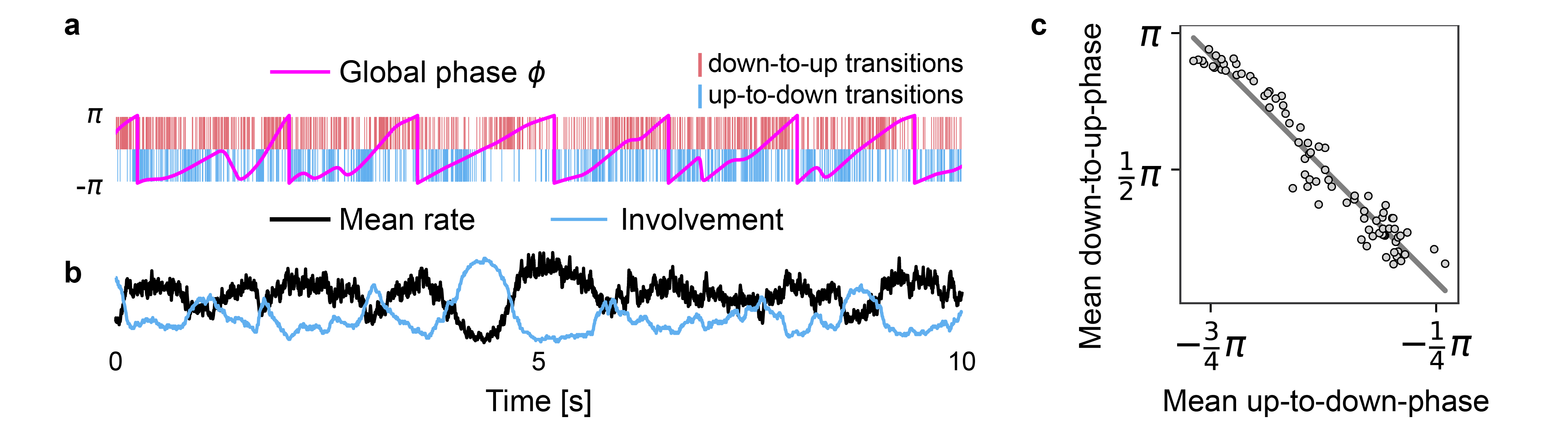}
	\caption{\textbf{Involvement time series defines the global oscillation phase of an event with respect to a cortical \textit{down-state}.}
		\textbf{(a)} Global phase $\phi$ (pink) defined by the phase of the involvement time series superimposed with \textit{down-state} (blue vertical lines) and \textit{up-state} transitions (red vertical lines) of all brain areas.
		\textbf{(b)} Time series of \textit{down-state} involvement (blue), measuring the fraction of brain areas in the \textit{down-state}, and of the mean firing rate across all areas (black).
		\textbf{(c)} Brain areas that initiate \textit{down-states} first, transition to \textit{up-states} last. The mean \textit{up-to-down} and \textit{down-to-up} transition phases for each node in the brain-network are plotted against each other. The linear regression line has a slope of $-1.76$ ($R^2 = 0.92$ and $p<0.001$). Results were obtained with the sleep model in Fig. \ref{fig:sleep} in the main manuscript.
	}
	\label{fig:supp:sleep_model}
	\end{figure}

	To determine the mean phase of \textit{up-} and \textit{down-state} transitions relative to the global (whole-brain) oscillation for every brain area (Suppl. Fig. \ref{fig:supp:sleep_model} a), we first compute the global phase $\phi(t)$ of SOs using the Hilbert transform of the \textit{down-state} involvement time series $I(t)$ whose inverse tightly tracks the mean firing rate of the brain (Suppl. Fig. \ref{fig:supp:sleep_model} b). 
	The signal $I(t)$ was first bandpass filtered between $0.5 \si{\hertz}$ and $2.0 \si{\hertz}$ using an implementation of the Butterworth filter \textit{scipy.signal.butter} of order 8 in SciPy. The signal was then converted into a comnplex-valued analytic signal $R(t)= A(t) \exp\big(i\phi(t)\big)$ using \textit{scipy.signal.hilbert\_transform}. The phase $\phi(t)$ of $R(t)$ then served as the the phase of whole-brain oscillations.
	
	\subsubsection*{Measure of synchrony and metastability}
\begin{figure}[H]
	\centering
	\includegraphics[width=\linewidth]{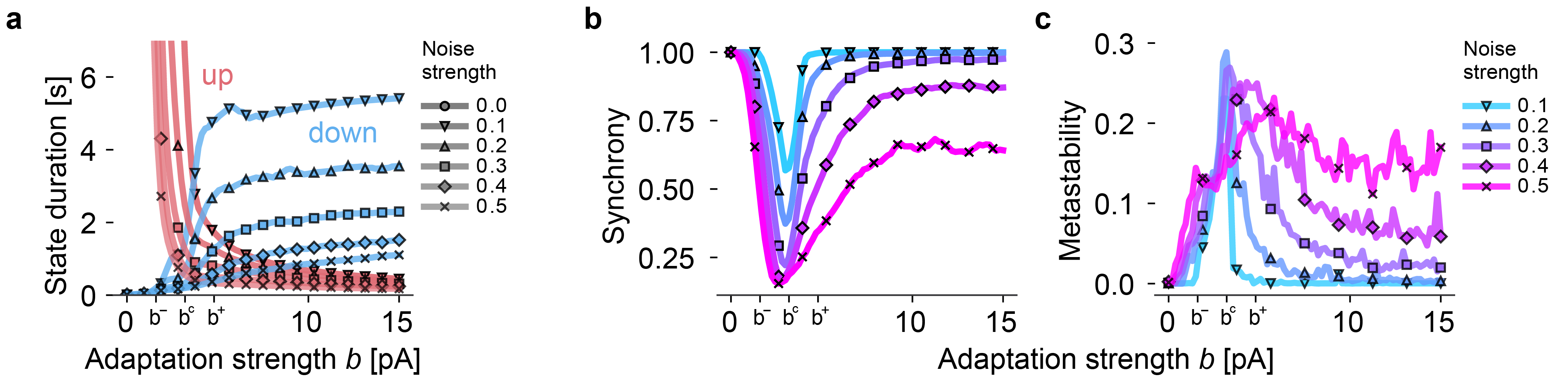}
	\caption{\textbf{The statistics of slow oscillations depends on adaptation strength.}
		\textbf{(a)} Average state durations for \textit{up-} (red) and \textit{down-states} (blue) as a function of the adaptation strength $b$ and the noise level $\sigma_\text{ou}$ (different lines). Noise strength is measured in $\si{\milli \volt}/\text{ms}^{3/2}$. 
		Tick marks $b^{-}$, $b^{c}$, and $b^{+}$ indicate the values for each panel in Fig. \ref{fig:adaptation} a in the main manuscript. $b^{c}$ denotes the best-fitting value of $b$ obtained during the optimization procedure used in Fig. \ref{fig:sleep} in the main manuscript. 
		\textbf{(b)} Synchrony of transitions to the \textit{down-state} as measured by the temporal mean of the Kuramoto order parameter $R(t)$ (see Eq. \ref{eq:Kuramoto} in the main manuscript).
		\textbf{(c)} Metastability defined as the temporal fluctuation of $R(t)$. 
		Other parameters are taken from the sleep model in Fig. \ref{fig:sleep} in the main manuscript.
	}
	\label{fig:supp:adaptation}
	\end{figure}

	In Fig. \ref{fig:supp:adaptation} b and c the synchrony and metastability of transitions to the \textit{down-state} were quantified using the Kuramoto order parameter \cite{Kuramoto2003} $R(t)$ which measures the synchrony of all brain areas. In Fig. \ref{fig:supp:adaptation} b, the temporal mean of $R(t)$, and in Fig. \ref{fig:supp:adaptation} c, the temporal standard deviation of $R(t)$ are plotted.
	The Kuramoto order parameter $R(t)$ is given by
	\begin{eqnarray}
		R(t) = \frac{1}{N} \left| \sum_{j=1}^{N} e^{i \varphi_j(t)}  \right|,
	\label{eq:Kuramoto}
	\end{eqnarray}
	where $N$ is the number of brain areas, and $\varphi_j(t)$ is the phase of the \textit{down-state} transitions of each area $j$. The phase of \textit{down-state} transitions of region (index $j$ ommited) can be defined as
	\begin{eqnarray}
	\varphi(t) = 2 \pi \frac{t - t_n}{t_n - t_{n-1}},
	\label{eq:Kuramoto:phases}
	\end{eqnarray}
	where $t_n$ is the time of the last transition and $t_{n-1}$ the time of the second to last transition \cite{Rosenblum2001}.

	\subsection*{Numerical simulations}
	All simulations, the parameter explorations, and the optimization framework including the evolutionary algorithm are implemented as a Python package in our whole-brain neural mass modeling framework \textit{neurolib} \cite{cakan2021a} which can be found at \url{https://github.com/neurolib-dev/neurolib}. The forward Euler method was used for the numerical integration with an integration time step of dt = $0.1 \si{\milli \second}$. The code for reproducing all presented figures can be found at \url{https://github.com/caglarcakan/sleeping_brain}.
	
	\section*{Acknowledgments}
	We would like to thank Dr. Nikola Jajcay and Dr. Michael Scholz for their contributions and the valuable exchange during this project. We would like to thank Melissa Skarmeta for their help on processing the EEG data. This work was funded by the Deutsche Forschungsgemeinschaft (DFG, German Research Foundation) – Project number 327654276 – SFB 1315.	
	
	\newpage 
	\section*{Supplementary Figures}

	\begin{figure}[H]
	\centering
	\includegraphics[width=\linewidth]{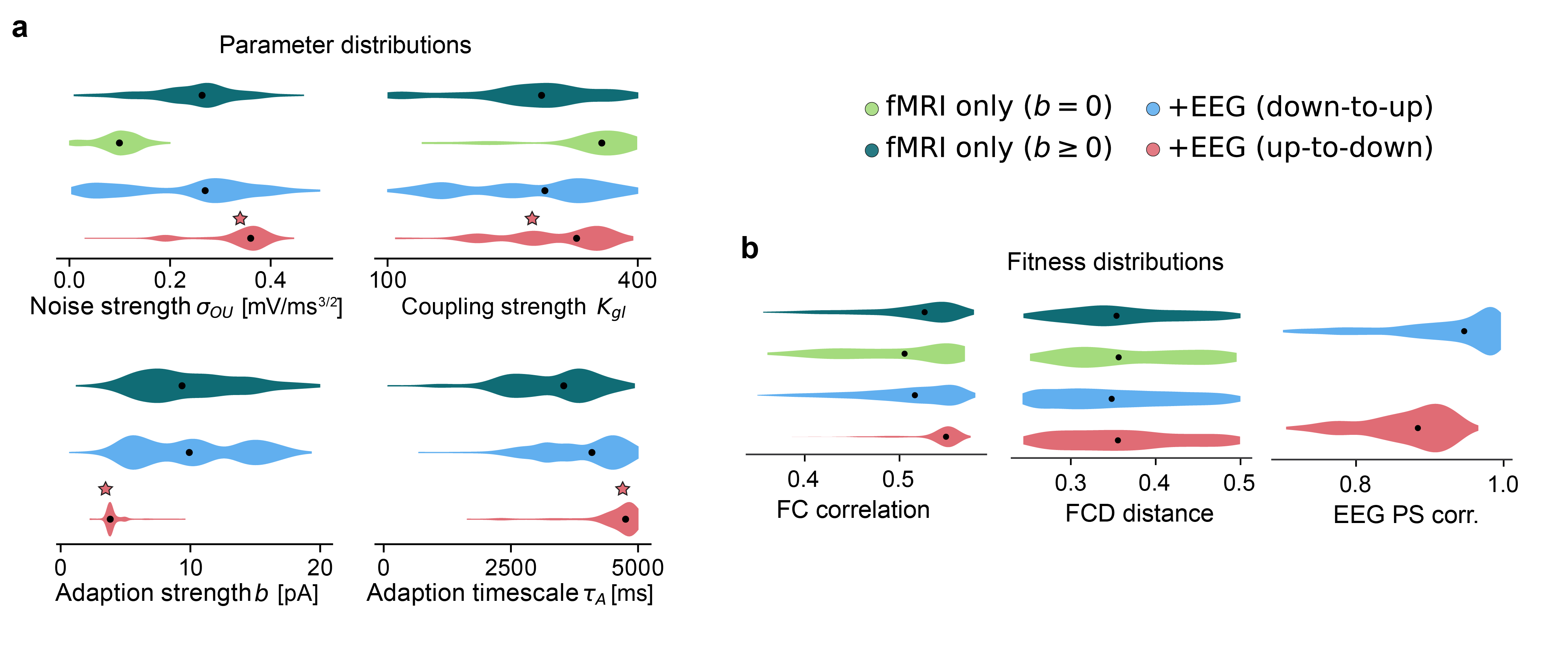}
	\caption{\textbf{Optimization results.}
	\textbf{(a)} Distributions of all all other optimized parameters across the final population of each optimization run additional to the input current strengths shown in Fig. \ref{fig:evolution} b in the main manuscript. Fit to fMRI data only without adaptation, $b=0 \si{\pico \ampere}$, in dark green and with adaptation, $b\geq0 \si{\pico \ampere}$, in light green. Fit to fMRI and EEG data with \textit{down-to-up} solutions (blue) and \textit{up-to-down} solutions (red). \textit{Up-to-down} fits (red) have stronger noise $\sigma_\text{ou}$ and weaker adaptation $b$ compared to \textit{down-to-up} fits (blue). Black dots show median values of the distributions. The star symbol indicates the parameters of the sleep model in Fig. \ref{fig:sleep} in the main manuscript.
	\textbf{(b)} Fitness distributions of the final population of each optimization.}
	\label{fig:supp:fits}
	\end{figure}

	\begin{figure}[H]
	\centering
	\includegraphics[width=\linewidth]{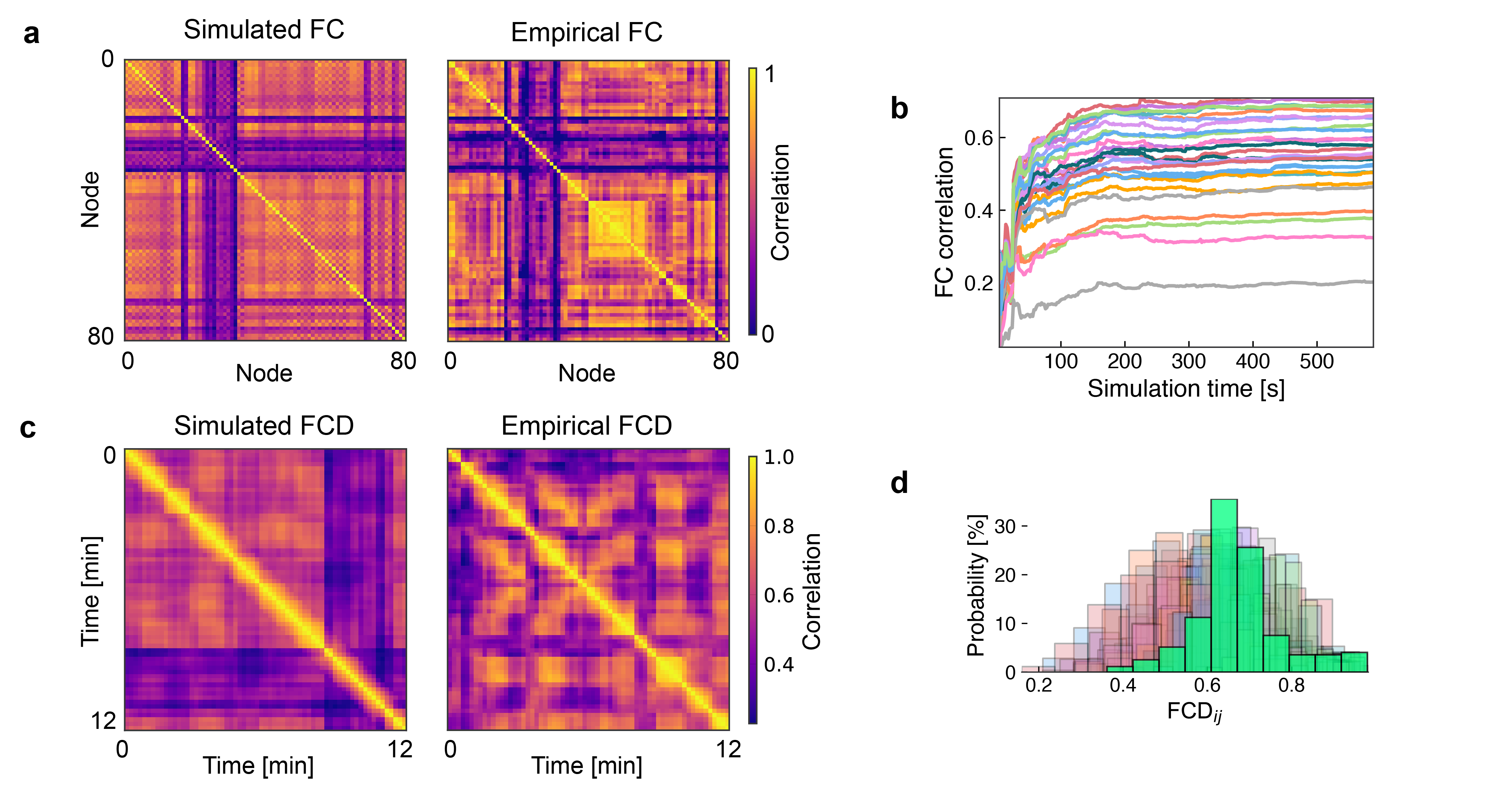}
	\caption{\textbf{Model fit to fMRI data.}
		\textbf{(a)} Simulated functional connectivity (FC) matrix of the sleep model and empirical FC matrix of the best fitting subject (from a total of 27 subjects). Color denotes the Pearson correlation coefficient of the BOLD time series for each pair of nodes in the brain network. Correlation between simulated and empirical matrices was $0.55$ averaged across all subjects with the best subject reaching $0.70$.
		\textbf{(b)} Correlation between simulated FC and all subjects FCs (different colors) as a function of the total simulation time.
		\textbf{(c)} Simulated and empirical functional connectivity dynamics (FCD) matrices. The empirical FCD matrix is shown for a the best-fitting subject. The Kolmogorov-Smirnoff distance between the distribution of the FCD matrix entries to the empirical FCD matrix entries averaged across all subjects was $0.28$ with the best subject reaching $0.07$. 
		\textbf{(d)} Distributions of lower diagonal entries of the simulated FCD matrix (solid green) and the empirical data for each subject (different colors). Parameters are taken from the sleep model in Fig. \ref{fig:sleep} in the main manuscript (star in Fig. \ref{fig:evolution} b in the main manuscript and in Suppl. Fig. \ref{fig:supp:fits} a).
	}
	\label{fig:supp:fit_fmri}
\end{figure}

	\begin{figure}[H]
	\centering
	\includegraphics[width=0.8\linewidth]{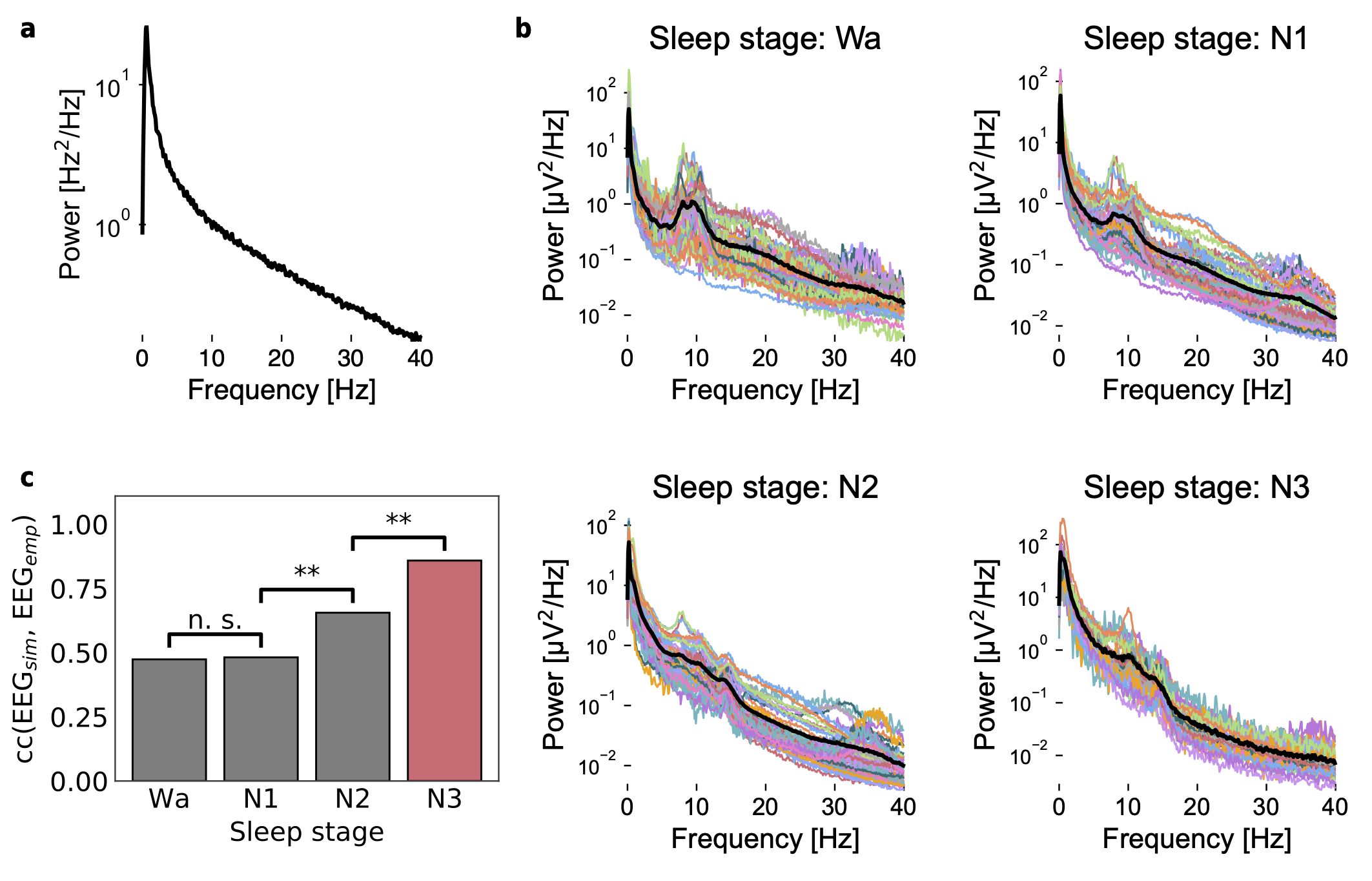}
	\caption{\textbf{Model fit to EEG data.}
		\textbf{(a)} Power spectrum of the mean firing rate of the sleep model (star in Fig. \ref{fig:evolution} b in the main manuscript and in Suppl. Fig. \ref{fig:supp:fits} a).
		\textbf{(b)} Empirical EEG power spectra of the awake state (Wa) and different sleep stages (N1-N3). Different colors denote the power spectra of the 18 different subjects. The black line denotes the across-subject average. 
		\textbf{(c)} The model fits best to the empirical EEG power spectra during sleep stage N3 which was used as the target for fitting the model. The bars show the subject-averaged Pearson correlation coefficient (cc) between the power spectrum of the sleep model (EEG$_{sim}$) and the empirical power spectra of different sleep stages (EEG$_{emp}$). Sleep stage N3 is indicated in red. Means and standard deviations of the correlation coefficients are: $0.47 \pm 0.11$ (Wa), $0.48 \pm 0.08$ (N1), $0.66 \pm 0.12$ (N2), $0.86 \pm 0.07$ (N3). Parameters are taken from the sleep model in Fig. \ref{fig:sleep} in the main manuscript.
	}
	\label{fig:supp:fit_eeg}
	\end{figure}
	
	\newpage 
	\bibliographystyle{apacite}
	\bibliography{refs.bib}

\end{document}